\documentclass[a4paper,12pt]{article}
\usepackage{latexsym}
\newcommand{\pair}[1]{\langle{#1}\rangle}
\newcommand{\BQED}{\hfill \hbox{\rule{8pt}{8pt}}}

\newenvironment{namelist}[1]{%
\begin{list}{}
  { 
	\settowidth{\labelwidth}{#1}
	\setlength{\leftmargin}{1.1\labelwidth}}
  \setlength{\itemsep}{0cm}
}{%
\end{list}}
\catcode`\@=11
\def\newexample#1{\@ifnextchar[{\@oexm{#1}}{\@nexm{#1}}}

\def\@nexm#1#2{%
\@ifnextchar[{\@xnexm{#1}{#2}}{\@ynexm{#1}{#2}}}

\def\@xnexm#1#2[#3]{\expandafter\@ifdefinable\csname #1\endcsname
{\@definecounter{#1}\@addtoreset{#1}{#3}%
\expandafter\xdef\csname the#1\endcsname{\expandafter\noexpand
  \csname the#3\endcsname \@exmcountersep \@exmcounter{#1}}%
\global\@namedef{#1}{\@exm{#1}{#2}}\global\@namedef{end#1}{\@endexample}}}

\def\@ynexm#1#2{\expandafter\@ifdefinable\csname #1\endcsname
{\@definecounter{#1}%
\expandafter\xdef\csname the#1\endcsname{\@exmcounter{#1}}%
\global\@namedef{#1}{\@exm{#1}{#2}}\global\@namedef{end#1}{\@endexample}}}

\def\@oexm#1[#2]#3{\expandafter\@ifdefinable\csname #1\endcsname
  {\global\@namedef{the#1}{\@nameuse{the#2}}%
\global\@namedef{#1}{\@exm{#2}{#3}}%
\global\@namedef{end#1}{\@endexample}}}

\def\@exm#1#2{\refstepcounter
    {#1}\@ifnextchar[{\@yexm{#1}{#2}}{\@xexm{#1}{#2}}}

\def\@xexm#1#2{\@beginexample{#2}{\csname the#1\endcsname}\ignorespaces}
\def\@yexm#1#2[#3]{\@opargbeginexample{#2}{\csname
       the#1\endcsname}{#3}\ignorespaces}

\def\@exmcounter#1{\noexpand\arabic{#1}}
\def\@exmcountersep{.}
\def\@beginexample#1#2{\trivlist \item[\hskip 
\labelsep{\bf #1\ #2:}]}
\def\@opargbeginexample#1#2#3{\trivlist
      \item[\hskip \labelsep{\bf #1\ #2\ }#3{\bf :}]}
\def\@endexample{\endtrivlist}

\catcode`\@=12

\newtheorem{lemma}{{\bf Lemma}}[section]
\newtheorem{thm}{{\bf Theorem}}[section]

\newtheorem{claim}{{\bf Claim}}[section]

\catcode`\@=11

\def\@xnthm#1#2[#3]{\expandafter\@ifdefinable\csname #1\endcsname
{\@definecounter{#1}\@addtoreset{#1}{#3}%
\expandafter\xdef\csname the#1\endcsname{\expandafter\noexpand
\bf \csname the#3\endcsname \@thmcountersep \@thmcounter{#1}}%
\global\@namedef{#1}{\@thm{#1}{#2}}\global\@namedef{end#1}{\@endtheorem}}}

\def\@ynthm#1#2{\expandafter\@ifdefinable\csname #1\endcsname
{\@definecounter{#1}%
\expandafter\xdef\csname the#1\endcsname{\bf \@thmcounter{#1}}%
\global\@namedef{#1}{\@thm{#1}{#2}}\global\@namedef{end#1}{\@endtheorem}}}

\def\@begintheorem#1#2{\trivlist \item[\hskip 
\labelsep{\bf #1~#2:}]\sl}
\def\@opargbegintheorem#1#2#3{\trivlist
      \item[\hskip \labelsep{\bf #1~#2}~#3{\bf :}]\sl}

\catcode`\@=12

\newbox\rubisita
\newbox\rubiue
\newdimen\rubiw
\def\rubi#1#2{{\setbox\rubisita=\hbox{#1}\setbox\rubiue=\hbox{\tiny #2}%
\ifdim \wd\rubisita>\wd\rubiue\rubiw=\wd\rubisita\else\rubiw=\wd\rubiue\fi%
\kanjiskip=0pt plus1fil%
\setbox\rubisita=\hbox to \rubiw{\hfil#1\hfil}%
\setbox\rubiue=\hbox to \rubiw{\tiny\hfil#2\hfil}%
\vbox{\offinterlineskip\box\rubiue\break\box\rubisita}}}
\begin{document}
%
%
%
%
%
\begin{center}
{\Large {\bf Greedy Algorithms for Multi-Queue}}\\
{\Large {\bf  Buffer Management with Class Segregation}}\bigskip\\
\begin{tabular}{ccc}
{\sc Toshiya Itoh} & & {\sc Seiji Yoshimoto}\\
{\sf titoh@ip.titech.ac.jp} & & {\sf yoshimoto.s.aa@m.titech.ac.jp}\\
Imaging Science \& Engineering Laboratory & & Department of Computer 
Science\\
Tokyo Institute of Technology &  & Tokyo Institute of Technology\\
\end{tabular}
\end{center}\medskip
{\bf Abstract:} In this paper, we focus on a multi-queue buffer management 
in which  packets~of~different values are segregated in different queues. 
Our model consists of $m$ packets values~and~$m$ queues. 
Recently, Al-Bawani and Souza (CoRR abs/1103.6049v2 [cs.DS]19 Sep 2011) 
presented an online multi-queue buffer management algorithm {\sf Greedy} 
and showed that~it~is~2-competi\-tive~for the general $m$-valued case, i.e., 
$m$ packet values are $0<v_{1}<v_{2}<\cdots<v_{m}$, 
and it is $(1+v_{1}/v_{2})$-competitive for the two-valued 
case, i.e., two packet values are $0<v_{1}<v_{2}$.
For the general $m$-valued case, let 
$c_{i} = (v_{i}+\sum_{j=1}^{i-1}2^{j-1}v_{i-j})/(v_{i+1}+\sum_{j=1}^{i-1}2^{j-1}v_{i-j})$ 
for $1 \leq i \leq m-1$,~and let $c_{m}^{*} = \max_{i} c_{i}$. 
In~this~paper, we precisely analyze 
the competitive ratio of {\sf Greedy}~for~the~general $m$-valued 
case, and 
show that the algorithm {\sf Greedy} is $(1+c_{m}^{*})$-competitive. \medskip
\section{Introduction} \label{sec-introduction} 
%
Due to the burst growth of the Internet use, network traffic has 
increased year by year.~This~overloads networking systems and 
degrades the quality of communications, e.g., loss of bandwidth, 
packet drops, delay of responses, etc. 
To overcome such degradation of the communication~quality, 
the notion of Quality of Service (QoS) 
has received attention in practice, and is implemented by assigning 
nonnegative numerical values to packets to provide them with differentiated 
levels of service (priority). Such a packet value corresponds to the 
predefined Class of Service~(CoS).~In 
general, switches have several number of queues and each 
queue has a {\it buffer\/} to store arriving packets. 
Since network traffic changes frequently, 
switches need to control arriving packets to maximize 
the total priorities of transmitted packets, which is called 
{\it buffer management\/}.~Basical\-ly, 
%
switches have no knowledge on the arrivals of packets 
in the future when it manages~to~control new packets arriving to the 
switches. 
So the decision made by buffer management algorithm~can 
be regarded as an {\it online algorithm\/}, and in general, 
the performance of online algorithms is measured by  
{\it competitive ratio\/} \cite{BE}. 
Online buffer management algorithms can be classified~into~two 
types of queue management  
(one~is~{\it preemptive\/}~and~the~other~is~{\it nonpreemptive\/}).  
Informally,~we say that an online bufffer management algorithm is 
preemptive if it is allowed~to~discard packets buffered 
in the queues on the arrival of 
new packets; nonpreemptive otherwise (i.e., all packets buffered in 
the queues will be eventually transmitted). 
%
\subsection{Multi-Queue Buffer Management} \label{subsec-buffer}
%
In this paper, we focus on a multi-queue model  in which packets 
of different values are segregated in different queues (see, e.g., \cite{CS}, \cite{LL}). 
Our model consists of $m$ packet values~and $m$ 
queues\footnote[1]{~In general, we can consider a model of $m$ packet 
values and $n$ queues (with $m \neq n$), but in this paper, we deal with 
only a model of $m$ packet values and $m$ queues.}. 
Let ${\cal V}=\{v_{1},v_{2},\ldots,v_{m}\}$ be the set 
of $m$ nonnegative {\it packet values\/}, where $0<v_{1}<v_{2}<\cdots<v_{m}$, 
and let ${\cal Q}=\{Q_{1},Q_{2},\ldots,Q_{m}\}$ be the set of $m$ queues. 
A packet of value $v_{i} \in  {\cal V}$~is~referred~to as a {\it $v_{i}$-packet\/}, 
and a queue storing $v_{i}$-packets is referred to as a $v_{i}$-queue. Without 
loss~of~gener\-ality, we assume 
that $Q_{i} \in {\cal Q}$ is a $v_{i}$-queue 
for each $i \in [1,m]$\footnote[2]{~For any pair of integers $a \leq b$, let $[a,b]= \{a,,a+1,\ldots,b\}$. }. Each~$Q_{i} \in {\cal Q}$~has~a~capacity~$B_{i}\geq 1$,  
i.e., each $Q_{i} \in {\cal Q}$ can store up to $B_{i} \geq 1$ 
packets. Since all packets buffered in queue~$Q_{i} \in {\cal Q}$ have 
the same value $v_{i} \in {\cal V}$, the order of 
transmitting packets is irrelevant. 

For convenience, we assume that time is discretized into slot of unit length. 
Packets arrive over time and each arriving packet is assigned with a unique 
(nonintegral) arrival time,~a~value $v_{i} \in {\cal V}$, and its destination queue 
$Q_{i} \in {\cal Q}$ (as we have assumed, $Q_{i} \in {\cal Q}$ 
is a $v_{i}$-queue).
We use $\sigma=\pair{e_{0},e_{1},e_{2},\ldots}$ to denote a sequence of 
{\it arrive events\/}  and {\it send events\/}, where an arrive event 
corresponds to the arrival of a new packet and a send event 
corresponds to the transmission of a packet buffered in queues at 
integral time (i.e., the end of time slot).~An~online~(multi-queue) buffer 
management algorithm {\sf Alg} consists of 
two phases: {\it admission\/} phase {\it schedulilng\/}~phases.~In 
the admission phase, {\sf Alg} must decide on the arrival 
of a packet whether~to~accept or reject~the packet 
without any knowledge on the 
future arrivals of packets (if {\sf Alg} is preemptive, 
then~it~may discard packets buffered in queues in the admission phase). 
In the scheduling phase, {\sf Alg} chooses 
one of the nonempty queues at send event and exactly one 
packet is transmitted~out~of~the~queue chosen. 
Since all packets buffered in the same queue have 
the same~value,~preemption does not make sense in our model. Thus a packet 
accepted must eventually be transmitted. 

We say that an (online and offline) 
algorithm is {\it diligent\/} if (1) it must accept a packet~arriving to its destination 
queue when the destination queue has vacancies, and 
(2) it must transmit a packet when it has nonempty queues. 
It is not difficult to see that any nondiligent algorithm~can be 
transformed to a diligent algorithm without decreasing its benefit 
(sum~of~values~of~transmitted packets). Thus in this paper, 
we focus on only diligent algorithms. 
%
%
%
\subsection{Main Results} \label{subsec-main}
%
Al-Bawani and Souza \cite{ABS} recently 
presented an online multi-queue buffer 
management algorithm {\sf Greedy} and showed that 
it is 2-competitive for the general $m$-valued case, i.e., $m$ packet values 
are $0<v_{1}<v_{2}<\cdots<v_{m}$, and 
$(1+v_{1}/v_{2})$-competitive~for~the~two-valued case, i.e., $m=2$. 

For the general $m$-valued case,  
let $c_{m}^{*}=\max_{i} c_{i}$, where for each $1 \leq i \leq m-1$, 
\[
c_{i} = \frac{v_{i}+\sum_{j=1}^{i-1}2^{j-1}v_{i-j}}{v_{i+1}+\sum_{j=1}^{i-1}2^{j-1}v_{i-j}}.
\]
In this paper, we precisely analyze the competitive ratio of {\sf Greedy} 
for the general $m$-valued~case, and 
show that the algorithm {\sf Greedy} is $(1+c_{m}^{*})$-competitive (see Theorem 
\ref{thm-main}). 
Note that $c_{m}^{*}<1$. Thus we have that $1+c_{m}^{*}<2$ and 
for the general $m$ valued 
case, our results improves the known result that the algorithm 
{\sf Greedy} is 2-competitive \cite[Theorem 2.1]{ABS}. 

For example, let 
us consider the case that $v_{1}=1,v_{2}=2$, and 
$v_{i+1}=v_{i}+\sum_{j=1}^{i-1}2^{j-1}v_{i-j}$~for each $i \in [2, m-1]$. 
It is obvious that $0<v_{1}<v_{2}<\cdots<v_{m}$ and 
$c_{m}^{*}=\max_{i}c_{i}=1/2$.~Thus for those packet values, our 
result guarantees that the algorithm {\sf Greedy} is 3/2-competitive, 
while the known result only guarantees that 
the algorithm {\sf Greedy} is 2-competitive \cite[Theorem 2.1]{ABS}.

%
\subsection{Related Works} \label{subsec-related}
%
The competitive analysis for the buffer management policies for switches 
were initiated~by~Aiello et al. \cite{AMRR}, Mansour et al. \cite{MPSL}, and 
Kesselman et al. \cite{Ketal}, and the extensive studies~have been made for several 
models (for comprehensive surveys, see, e.g., 
\cite{A},\cite{ES},\cite{J},\cite{C},\cite{G}). 

The model we deal with in this paper can be regarded as the generalization of 
unit-valued model, where  the switches consist of $m$ queues 
of the same buffer size $B$ and all packets have~unit value, i.e., 
$v_{1}=v_{2}=\cdots=v_{m}$. The following tables summarize the known results: 
\begin{table*}[htb]
\caption{Deterministic Competitive Ratio 
(Unit-Valued Multi-Queue Model)} \label{tab-det-cr}
\begin{center}
\def\arraystretch{1.2}
\begin{tabular}{|c|c||c|c|} \hline
\multicolumn{2}{|c||}{\makebox[5.0cm][c]{Upper Bound}} & 
\multicolumn{2}{c|}{\makebox[5.0cm][c]{Lower Bound}}\\  \hline
\begin{tabular}{cr}
\makebox[2.5cm][c]{2} & \makebox[1.0cm][r]{\cite{AR}}\\
\makebox[2.5cm][c]{1.889}  & \makebox[1.0cm][r]{\cite{AS}}\\
\makebox[2.5cm][c]{1.857} & \makebox[1.0cm][r]{\cite{AS}}\\
\makebox[2.5cm][c]{$\frac{e}{e-1}\approx 1.582$} & 
\makebox[1.0cm][r]{\cite{AL}}
\end{tabular} & 
\begin{tabular}{c}
---\\
\makebox[1.5cm][c]{$m \gg B$}\\
\makebox[1.5cm][c]{$B=2$}\\
\makebox[1.5cm][c]{large $B$}
\end{tabular} & 
\begin{tabular}{cr}
\makebox[2.5cm][c]{$2-1/m$} & \makebox[1.0cm][r]{\cite{AR}}\\
\makebox[2.5cm][c]{$1.366-\Theta(1/m)$} & \makebox[1.0cm][r]{\cite{AR}}\\
\makebox[2.5cm][c]{$\frac{e}{e-1}\approx 1.582$} & 
\makebox[1.0cm][r]{\cite{AS}}
\end{tabular} & 
\begin{tabular}{c}
\makebox[1.5cm][c]{$B=1$}\\
\makebox[1.5cm][c]{$B\geq 1$}\\
---
\end{tabular}\\ \hline
\end{tabular}
\end{center}
%
%
%
\caption{Randomized Competitive Ratio 
(Unit-Valued Multi-Queue Model)} \label{tab-rand-cr}
\begin{center}
\def\arraystretch{1.2}
\begin{tabular}{|c|c||c|c|} \hline
\multicolumn{2}{|c||}{\makebox[5.0cm][c]{Upper Bound}} & 
\multicolumn{2}{c|}{\makebox[5.0cm][c]{Lower Bound}}\\  \hline
\begin{tabular}{cr}
\makebox[2.5cm][c]{$\frac{e}{e-1}\approx 1.582$} & 
\makebox[1.0cm][r]{\cite{AR}}\\
\makebox[2.5cm][c]{1.231} & \makebox[1.0cm][r]{\cite{BM}}
\end{tabular} & 
\begin{tabular}{c}
\makebox[1.5cm][c]{$B > \log m$}\\
\makebox[1.5cm][c]{$m=2$}
\end{tabular} & 
\begin{tabular}{cr}
\makebox[2.5cm][c]{$1.46-\Theta(1/m)$}  & 
\makebox[1.0cm][r]{\cite{AR}}\\
\makebox[2.5cm][c]{1.466}  & \makebox[1.0cm][r]{\cite{AS}}\\
\makebox[2.5cm][c]{1.231}  & \makebox[1.0cm][r]{\cite{AS}}
\end{tabular} & 
\begin{tabular}{c}
\makebox[1.5cm][c]{$B=1$}\\
\makebox[1.5cm][c]{large $m$}\\
\makebox[1.5cm][c]{$m=2$}
\end{tabular}\\ \hline
\end{tabular}
\end{center}
\end{table*}

On the other hand, the model we deal with in this paper can be regarded as 
a special~case~of 
the general-valued multi-queue model where each of $m$ FIFO queues 
can buffer at most~$B$~packets of different values. 
For the preemptive multi-queue buffer management,~Azar~and~Richter~\cite{AR} presented a  
$(4+2\ln \alpha)$-competitive algorithm 
for the general-valued case (packet values~lie between 1 and $\alpha$) 
and a 2.6-competitive algorithm for the two-valued case 
(packet values~are~$v_{1}<v_{2}$, where $v_{1}=1$ and $v_{2}=\alpha$). 
For the general-valued case, Azar and Righter~\cite{AR2}~proposed~a more  
efficient algorithm {\sc transmit-largest head} ({\sc tlh}) that is 3-competitive, 
which is shown to be $(3-1/\alpha)$-competitive by Itoh and Takahashi \cite{IT}. 
%
\section{Preliminaries} \label{sec-preliminary}
%
For a sequence $\sigma'$ of arriving packets, we use 
$\sigma=\pair{e_{0},e_{1},e_{2},\ldots}$ to denote a sequence 
of~arrive and send events. Notice that 
an arrive event corresponds to the arrival of a new packet~(at~nonintegral time) 
and a send event corresponds to the transmission of a packet 
buffered in queues~at integral time. 
The online algorithm {\sf Greedy} works as follows: At send event, 
{\sf Greedy} transmits~a 
packet from the nonempty queue with highest packet 
value\footnote[3]{~Since $Q_{i} \in {\cal Q}$ is a $v_{i}$-queue, such a 
nonempty queue with highest packet value is {\it unique\/} if it exists.}, i.e., 
{\sf Greedy} transmits a $v_{h}$-packet if $v_{h}$-queue~is~non\-empty and all  
$v_{\ell}$-queues are empty for $\ell \in [h+1,m]$.
At arrive event, {\sf Greedy} accepts packets~in its 
destination queue until the corresponding queue becomes full. 

For an {\it online\/} algorithm {\sf Alg} and a sequence $\sigma$ 
of arrive and send events, 
we use ${\sf Alg}(\sigma)$ to denote the {\it benefit\/} of the algorithm 
{\sf Alg} on the sequence $\sigma$, i.e., 
the sum of values~of~packets~transmitted by the algorithms {\sf Alg}  
on the sequence $\sigma$. 
For a sequence $\sigma$ of arrive and send~events,~we 
also use ${\sf Opt}(\sigma)$ to denote 
the {\it benefit\/} of~the {\it optimal offline\/} algorithm {\sf Opt} 
on~the~sequence~$\sigma$,~i.e., 
the sum of values of packets transmitted by the 
optimal offline algorithm {\sf Opt} that knows the entire 
sequence $\sigma$ in advance. 
Our goal is to design an efficient (deterministic) online algorithm 
{\sf Alg} that minimizes 
${\sf Opt}(\sigma)/{\sf Alg}(\sigma)$~for~any sequence $\sigma$. 

At event $e_{i}$, let $A_{h}(e_{i})$  and $A_{h}^{*}(e_{i})$   
be the total  number of $v_{h}$-packets accepted by 
{\sf Greedy}~and {\sf Opt} until the event $e_{i}$, respectively,
$\delta_{h}(e_{i})$ and 
$\delta_{h}^{*}(e_{i})$ be the total number of~$v_{h}$-packets 
transmitted by {\sf Greedy} and {\sf Opt} until the event $e_{i}$, respectively, 
and $q_{h}(e_{i})$ and $q_{h}^{*}(e_{i})$ be the total number 
of $v_{h}$-packets 
buffered in $v_{h}$-queue of {\sf Greedy} and {\sf Opt} just after 
the~event~$e_{i}$,~respectively.~It~is immediate to see that 
for each $h \in [1,m]$ and each event $e_{i}$, 
\begin{eqnarray}
A_{h}(e_{i}) & =& \delta_{h}(e_{i})+q_{h}(e_{i}); \label{eq-preserve-greedy}\\
A_{h}^{*}(e_{i}) & = & \delta_{h}^{*}(e_{i})+q_{h}^{*}(e_{i}). 
\label{eq-preserve-opt}
\end{eqnarray}
For a sequence $\sigma$, let $A_{h}(\sigma)$ and 
$A_{h}^{*}(\sigma)$ be the total number of $v_{h}$-packets accepted by 
{\sf Greedy}~and {\sf Opt} until the end of the sequence $\sigma$, respectively, 
$\delta_{h}(\sigma)$ and $\delta_{h}^{*}(\sigma)$ be 
the total number~of~$v_{h}$-pack\-ets 
transmitted by {\sf Greedy} and {\sf Opt} 
until the end of the sequence $\sigma$, respectively, 
and~$q_{h}(\sigma)$~and $q_{h}^{*}(\sigma)$ be the number 
of $v_{h}$-packets buffered in $v_{h}$-queue of {\sf Greedy} and {\sf Opt} 
at the end~of~the~sequence $\sigma$, respectively. It is immediate to see that 
$q_{h}(\sigma)=q_{h}^{*}(\sigma)=0$ for each $h \in [1,m]$.~So~from 
Eqs. (\ref{eq-preserve-greedy}) and (\ref{eq-preserve-opt}), 
it follows that $A_{h}(\sigma)=\delta_{h}(\sigma)$ and 
$A_{h}^{*}(\sigma)=\delta_{h}^{*}(\sigma)$ for each $h \in [1,m]$. 

For the general $m$-valued case, 
Al-Bawani and Souza showed the following result~on~the~number of 
packets accepted by {\sf Greedy} and {\sf Opt}, 
which is crucial in the subsequent discussions. 
\begin{lemma}[\mbox{\cite[Lemma 2.2]{ABS}}]  \label{lemma-AS}
For each $h \in [1,m]$, the following holds$:$
\[
\sum_{\ell=h}^{m} \left \{ A_{\ell}^{*}(\sigma) - A_{\ell}(\sigma) \right\}
\leq \sum_{\ell=h}^{m} A_{\ell}(\sigma). 
\]
\end{lemma} 

Assume that in the sequence $\sigma=\pair{e_{0},e_{1},e_{2},\ldots}$, 
there exist $k \geq 1$ send events, and for each $j \in [0,k]$, 
let $s_{j}$ be the $j$th send event, where $s_{0}=e_{0}$
is an initial send event that transmits~a~null packet. 
For each $j \in [1,k]$, we use $\Sigma_{j}$ to denote the set  
of arrive events between send event $s_{j-1}$ and send event $s_{j}$, i.e., 
$\Sigma_{j}$ consists of arrive events after send event $s_{j-1}$ and 
before send event $s_{j}$. Notice that $\Sigma_{j}$ could be an empty set. 
%
%
%
\section{Relationships Between Greedy and Opt} \label{sec-greedy-vs-opt}
%
\subsection{Number of Transmitted Packets} \label{subsec-transmitted}
%
In this subsection, we investigate the relationships between 
the number of packets transmitted by {\sf Greedy} and 
the number of packets transmitted by {\sf Opt}. 
%
For each $h \in [1,m-1]$~and~each~event $e_{i}$, let $\xi_{h}(e_{i})=
\delta_{h}(e_{i})+\cdots + \delta_{m}(e_{i})-\delta_{h}^{*}(e_{i})$. 
\begin{claim} \label{claim-1}
For each $h \in  [1,m-1]$ and 
each $j \in [2,k]$, if $q_{h}(s_{j-1})+\cdots + q_{m}(s_{j-1}) > 0$   
$($i.e.,~just after $s_{j-1}$, a 
nonempty $v_{\ell}$-queue of {\sf Greedy} with $\ell \in [h,m]$~exists$)$,~then~$\xi_{h}(s_{j})\geq \xi_{h}(s_{j-1})$. 
\end{claim}
{\bf Proof:} Since every $e_{i} \in \Sigma_{j}$ 
is arrive event, we have that for each $g \in [h,m]$, 
the~number~of~pack\-ets buffered in $v_{g}$-queue 
does not decrease at each arrive event $e_{i} \in \Sigma_{j}$. 
Then~from~the~assumption that $q_{h}(s_{j-1})+\cdots + q_{m}(s_{j-1}) > 0$, 
it follows that there exists an $\ell \in [h,m]$ 
such~that~$v_{\ell}$-queue of {\sf Greedy} is nonempty 
just before send event $s_{j}$. 
Thus from the definition of {\sf Greedy},~it~is~immedi\-ate to see that 
for some $r \in [\ell,m]$, {\sf Greedy} transmits a $v_{r}$-packet 
at send event $s_{j}$, which implies that $\delta_{h}(s_{j})+\cdots +\delta_{m}(s_{j})
=\delta_{h}(s_{j-1})+\cdots+\delta_{m}(s_{j-1})+1$. 
So we have that 
\begin{eqnarray*}
\xi_{h}(s_{j}) & = & \delta_{h}(s_{j})+\cdots+\delta_{m}(s_{j})-\delta_{h}^{*}(s_{j})\\
& \geq & \left\{\delta_{h}(s_{j-1})+\cdots+\delta_{m}(s_{j-1})+1\right\}-
\left\{\delta_{h}^{*}(s_{j-1})+1\right\}\\
& = & \delta_{h}(s_{j-1})+\cdots+\delta_{m}(s_{j-1})-\delta_{h}^{*}(s_{j-1})
= \xi_{h}(s_{j-1}),
\end{eqnarray*}
where the inequality follows from the fact that 
$\delta_{h}^{*}(s_{j}) \leq \delta_{h}^{*}(s_{j-1})+1$. \BQED
\begin{claim} \label{claim-3}
For each $h \in [1,m-1]$ and each $j \in [1,k]$, if 
$q_{h}^{*}(s_{j-1})=0$~$($i.e.,~just~after 
$s_{j-1}$,~
$v_{h}$-queue of {\sf Opt} is empty$)$, then 
$\xi_{h}(s_{j}) \geq \xi_{h}(s_{j-1})$. 
\end{claim}
{\bf Proof:} Let us consider the following cases: (1) 
$v_{h}$-queue of {\sf Opt} is 
empty just before send~event $s_{j}$
and (2) $v_{h}$-queue of {\sf Opt} is 
nonempty just before send event $s_{j}$. 
For the case (1), it~is~immedi\-ate to 
see that $\delta_{h}^{*}(s_{j})=\delta_{h}^{*}(s_{j-1})$. 
So we have that 
\begin{eqnarray*}
\xi_{h}(s_{j}) & = & \delta_{h}(s_{j})+\cdots+\delta_{m}(s_{j})-\delta_{h}^{*}(s_{j})\\
& \geq & \delta_{h}(s_{j-1})+\cdots+\delta_{m}(s_{j-1})-\delta_{h}^{*}(s_{j-1})
=\xi_{h}(s_{j-1}),
\end{eqnarray*}
where the inequality follows from the fact that 
$\delta_{h}(s_{j})+\cdots+\delta_{m}(s_{j})\geq \delta_{h}(s_{j-1})
+\cdots+\delta_{m}(s_{j-1})$.~For the case (2), 
there exists arrive event $e_{i} \in \Sigma_{j}$ such that a $v_{h}$-packet arrives, 
because~of~the~assumption that $v_{h}$-queue of {\sf Opt} 
is empty just after the send event $s_{j-1}$. 
Then~from~the~definition of {\sf Greedy}, it is easy to see that 
$v_{h}$-queue of {\sf Greedy} is nonemnty just before 
send~event~$s_{j}$~and~that at send event $s_{j}$, 
{\sf Greedy} transmits a $v_{\ell}$-packet with 
$\ell \in [h,m]$. 
This implies that $\delta_{h}(s_{j})+\cdots+
\delta_{m}(s_{j})= \delta_{h}(s_{j-1})+\cdots+\delta_{m}(s_{j-1})+1$. 
Thus it follows that 
\begin{eqnarray*}
\xi_{h}(s_{j}) & = & \delta_{h}(s_{j})+\cdots+\delta_{m}(s_{j})-\delta_{h}^{*}(s_{j})\\
& \geq & \left\{\delta_{h}(s_{j-1})+\cdots+\delta_{m}(s_{j-1})+1\right\}-
\left\{\delta_{h}^{*}(s_{j-1})+1\right\}\\
& = & \delta_{h}(s_{j-1})+\cdots+\delta_{m}(s_{j-1})-\delta_{h}^{*}(s_{j-1})
= \xi_{h}(s_{j-1}),
\end{eqnarray*}
where the inequality follows from the fact that $\delta_{h}^{*}(s_{j}) 
\leq \delta_{h}^{*}(s_{j-1})+1$. \BQED
\begin{lemma} \label{lemma-send}
For each $h \in [1,m-1]$ and each event $e_{i}$, 
$\xi_{h}(e_{i}) \geq 0$. 
\end{lemma}
{\bf Proof:} We show the lemma by induction~on~events $e_{i}$. 
It is obvious that $\xi_{h}(e_{0})=0$.~For~$t \geq 1$, assume 
that $\xi_{h}(e_{i})\geq 0$ for 
each $i \in [0,t-1]$. If $e_{t}$ is arrive event, then 
$\delta_{\ell}(e_{t})=\delta_{\ell}(e_{t-1})$~for~each $\ell \in [h,m]$ 
and 
$\delta_{h}^{*}(e_{t})=\delta_{h}^{*}(e_{t-1})$. This 
implies that $\xi_{h}(e_{t})=\xi_{h}(e_{t-1})$ and 
from~the~induction~hypothesis, it follows that 
$\xi_{h}(e_{t})=\xi_{h}(e_{t-1})\geq 0$. 
Thus in the rest of the proof,~we~focus~on~only send events and 
show the lemma by induction on send events $s_{j}$. 

{\sf Base Step:} We show that $\xi_{h}(s_{1})\geq 0$ at the first send 
event $s_{1}$. 
Let us consider the following cases: (1) there exists arrive event 
$e_{t} \in \Sigma_{1}$ at which a $v_{\ell}$-packet with $\ell \in [h,m]$ arrives 
and~(2) there exists no arrive event $e_{t} \in \Sigma_{1}$ at which 
a $v_{\ell}$-packet with $\ell \in [h,m]$ arrives.~For~the~case (1), 
we have that $v_{\ell}$-queue of {\sf Greedy} is nonempty just before 
send event $s_{1}$. So from~the~definition of 
{\sf Greedy}, it follows that $\delta_{h}(s_{1})+\cdots+\delta_{m}(s_{1})= 1$. 
Since $\delta_{h}^{*}(s_{1})\leq 1$, 
this~implies~that~$\xi_{h}(s_{1}) = \delta_{h}(s_{1}) + \cdots + \delta_{m}(s_{1}) 
- \delta_{h}^{*}(s_{1}) \geq 1 - 1 = 0$. 
For the case (2), 
it is immediate that 
$\delta_{h}(s_{1})=\cdots=\delta_{m}(s_{1})=0$ 
and $\delta_{h}^{*}(s_{1})=0$. 
Thus we have that $\xi_{h}(s_{1}) = \delta_{h}(s_{1}) + \cdots+\delta_{m}(s_{1}) 
- \delta_{h}^{*}(s_{1}) = 0-0=0$. 

{\sf Induction Step:} For $t \in [2,k]$, assume that $\xi_{h}(s_{j})\geq 0$ 
for each $j \in [0,t-1]$. Since~$\delta_{h}(s_{t})+\cdots+\delta_{m}(s_{t})\geq 
\delta_{h}(s_{t-1})+\cdots+\delta_{m}(s_{t-1})$ and 
$\delta_{h}^{*}(s_{t}) \leq \delta_{h}^{*}(s_{t-1})+1$, 
we have that if $\xi_{h}(s_{t-1}) \geq 1$,~then 
\begin{eqnarray*}
\xi_{h}(s_{t}) & = & \delta_{h}(s_{t})+\cdots+\delta_{m}(s_{t})-
\delta_{h}^{*}(s_{t})\\
& \geq & \delta_{h}(s_{t-1})+\cdots+\delta_{m}(s_{t-1})-\left\{
\delta_{h}^{*}(s_{t-1})+1\right\}\\
& = & \delta_{h}(s_{t-1})+\cdots+\delta_{m}(s_{t-1})-
\delta_{h}^{*}(s_{t-1})-1 = \xi_{h}(s_{t-1})-1\geq 0. 
\end{eqnarray*}
Thus we assume that $\xi_{h}(s_{t-1})=\delta_{h}(s_{t-1})
+\cdots+\delta_{m}(s_{t-1})-\delta_{h}^{*}(s_{t-1})=0$.
If $\delta_{h}^{*}(s_{t-1})=0$,~then~we have 
$\delta_{h}(s_{t-1})=\cdots=\delta_{m}(s_{t-1})=0$. 
From the definition of {\sf Greedy}, it follows that 
for~each~$\ell \in [h,m]$, no $v_{\ell}$-packets 
arrive until send event $s_{t-1}$, 
which implies that $q_{h}^{*}(s_{t-1})=0$.~So 
from Claim \ref{claim-3} and the induction hypothesis, it follows that 
$\xi_{h}(s_{t})\geq \xi_{h}(s_{t-1}) \geq 0$. 

Assume that $\delta_{h}^{*}(s_{t-1})=n>0$ 
and we consider the following cases: 
(3) {\sf Greedy}~does~not~reject any $v_{h}$-packet that arrives until 
send event $s_{t-1}$; 
(4) {\sf Greedy} rejects $v_{h}$-packets that arrive~until send event $s_{t-1}$. 
For the case (3), let $n_{h}$ be the number of $v_{h}$-packets that arrive 
until send~event $s_{t-1}$. It is obvious that 
$n_{h} \geq \delta_{h}^{*}(s_{t-1})=n >0$. 
If $q_{h}(s_{t-1}) > 0$,~then~from~Claim~\ref{claim-1} 
and~the~in\-duction hypothesis, it follows that $\xi_{h}(s_{t})
\geq \xi_{h}(s_{t-1})\geq 0$. 
Assume that $q_{h}(s_{t-1})=0$.~Since~$n_{h}>0$ $v_{h}$-packets 
arrive until send event $s_{t-1}$, $q_{h}(s_{t-1})=0$, and 
{\sf Greedy} does not reject 
any $v_{h}$-packet that arrives until send event $s_{t-1}$, 
we have that $\delta_{h}(s_{t-1})=n_{h}$. 
If $\delta_{h}^{*}(s_{t-1})< n_{h}$,~then~$\delta_{h}^{*}(s_{t-1})< n_{h} \leq \delta_{h}(s_{t-1})+\cdots+\delta_{m}(s_{t-1})$, 
which contradiction the assumption that $\delta_{h}(s_{t-1})+
\cdots+\delta_{m}(s_{t-1})-\delta_{h}^{*}(s_{t-1})=0$. 
So we assume that 
$\delta_{h}^{*}(s_{t-1})=n_{h}$. 
From~Eq. (\ref{eq-preserve-opt}) and the fact that 
$n_{h} \geq A_{h}^{*}(s_{t-1})$,~it  
is immediate that 
$q_{h}^{*}(s_{t-1}) = A_{h}^{*}(s_{t-1}) 
- \delta_{h}^{*}(s_{t-1})\leq n_{h}-n_{h}=0$, i.e., 
$q_{h}^{*}(s_{t-1})=0$.~So~from~Claim \ref{claim-3} and 
the induction hypothesis, it follows that $\xi_{h}(s_{t})\geq\xi_{h}(s_{t-1})\geq 0$. 

For the case (4), consider the following subcases: 
(4.1) $q_{h}(s_{t-1})>0$;~(4.2) $q_{h}(s_{t-1})=0$.~For the subcase (4.1), 
it is obvious that $q_{h}(s_{t-1})+\cdots+q_{m}(s_{t-1})>0$. 
Thus from~Claim~\ref{claim-1}~and~the~induction hypothesis, it follows 
that $\xi_{h}(s_{t})\geq \xi_{h}(s_{t-1}) \geq 0$. 
For the subcase (4.2), let $e_{\tau}$~be~the~last arrive event at which 
a $v_{h}$-packet is rejected by {\sf Greedy}. 
Assume that $e_{\tau} \in \Sigma_{j}$~for~some~$j \in [1,t-1]$, i.e., 
$e_{\tau}$ is arrive event between send event $s_{j-1}$ and send event $s_{j}$. 
Notice~that~the~$v_{h}$-queue of {\sf Greedy}~is~{\it full\/} just before arrive event 
$e_{\tau}$. This implies that $v_{h}$-queue of 
{\sf Greedy}~is~{\it full\/}~just before send event $s_{j}$. 
Let $L_{h} \geq 0$ be the total number of $v_{h}$-packets that arrive 
between~send events~$s_{j}$ and $s_{t-1}$. Since $q_{h}(s_{t-1})=0$, 
{\sf Greedy} must transmit $B_{h}+L_{h}$ 
$v_{h}$-packets from~send~event~$s_{j}$~to~send event $s_{t-1}$. So it 
follows that $\delta_{h}(s_{t-1})+\cdots+\delta_{m}(s_{t-1})\geq 
\delta_{h}(s_{j-1})+\cdots+\delta_{m}(s_{j-1})+B_{h}+L_{h}$.~Assume 
that {\sf Opt} transmits $K_{h}\geq 0$ $v_{h}$-packets 
at send events $s_{j},\ldots,s_{t-1}$,~i.e., 
$\delta_{h}^{*}(s_{t-1})= \delta_{h}^{*}(s_{j-1})+K_{h}$.~From~the induction 
hypothesis that $\xi_{h}(s_{j-1})\geq 0$, it follows that 
\begin{eqnarray*}
\xi_{h}(s_{t-1})& = & \delta_{h}(s_{t-1})+\cdots+
\delta_{m}(s_{t-1})-\delta_{h}^{*}(s_{t-1})\\
& \geq & \left\{ \delta_{h}(s_{j-1})+\dots+\delta_{m}(s_{j-1})+B_{h}+L_{h} \right\}
- \left\{\delta_{h}^{*}(s_{j-1})+K_{h} \right\}\\
& = & \delta_{h}(s_{j-1})+\cdots+\delta_{m}(s_{j-1}) - \delta_{h}^{*}(s_{j-1})
+B_{h}+L_{h} -K_{h}\\
&  = & \xi_{h}(s_{j-1}) + B_{h}+L_{h}-K_{h} \geq B_{h}+L_{h}-K_{h}.
\end{eqnarray*}
Note that $K_{h} \leq B_{h}+L_{h}$. 
If $K_{h} < B_{h}+L_{h}$, then it is immediate that 
$\xi_{h}(s_{t-1})>0$,~which~contradicts the assumption 
that $\xi_{h}(s_{t-1})=0$. So we have $K_{h}=B_{h}+L_{h}$, which implies that 
$q_{h}^{*}(s_{t-1})=0$.~Thus~from Claim \ref{claim-3} and 
the induction hypothesis, 
it follows that $\xi_{h}(s_{t})\geq \xi_{h}(s_{t-1})= 0$. \BQED
%
\subsection{Number of Accepted Packets} \label{subsec-accepted}
%
In this subsection, we investigate the relationships between 
the number of packets accepted by {\sf Greedy} and 
the number of packets accepted by {\sf Opt}. 
In the rest of this paper, we use $A_{h}$~and~$A^{*}_{h}$ 
instead of $A_{h}(\sigma)$ and $A^{*}_{h}(\sigma)$ respectively, 
when $\sigma$ is clear from the context. For each $h \in [1,m]$,  
let $D_{h}=A^{*}_{h}-A_{h}$ and $S_{h}=A_{h}+A_{h+1}+\cdots +A_{m}$. 

The following lemma shows the relationship between 
the number of $v_{m}$-packets accepted by {\sf Greedy} and 
the number of $v_{m}$-packets accepted by {\sf Opt}, which is 
a straightforward generalization of the result due to 
Al-Bawani~and~Souza \cite[Lemma 2.5]{ABS}.
\begin{lemma} \label{prop-acc3-eq}
%
$A_{m}=A_{m}^{*}$. 
\end{lemma}
{\bf Proof:} By definition of {\sf Greedy}, $v_{m}$-packet has 
priority at send event. Thus at any event~$e_{i}$,~the number of 
$v_{m}$-packets transmitted by {\sf Greedy} is maximum, i.e., 
$A_{m}(e_{i}) \geq A_{m}^{*}(e_{i})$. 

Assume that at arrive event $e_{t}$, 
$A_{m}(e_{t})$ becomes greater than $A_{m}^{*}(e_{t})$ 
for the first time, which implies that 
at arrive event $e_{t}$, {\sf Opt} rejects a $v_{m}$-packet but {\sf Greedy} 
accepts a $v_{m}$-packet.~Thus~just before 
event $e_{t}$, 
$v_{m}$-queue of {\sf Opt} is full but $v_{m}$-queue of {\sf Greedy} has 
at least one vacancy.~Since $A_{m}(e_{t-1})=A_{m}^{*}(e_{t-1})$, there must 
exist send event $e_{\tau}$ (with $\tau \leq t-1$) 
at~which~{\sf Opt}~transmitted~a $v_{\ell}$-packet with $\ell \in [1,m-1]$, 
while the $v_{m}$-queue of {\sf Opt} was not empty. 
Change~the~behavior~of {\sf Opt} at send event $e_{\tau}$ 
by transmitting a $v_{m}$-packet 
instead of the $v_{\ell}$-packet.~This~yields~an~increase in 
the benefit of {\sf Opt} and the $v_{m}$-packet 
rejected~at arrive event $e_{t}$ can be accepted. \BQED\medskip

The following lemma 
is a straightforward extension of the result by Al-Bawani and 
Souza~\cite[Lemma 2.6]{ABS} and plays a crucial role 
in the subsequent discussions. 
\begin{lemma} \label{prop-2}
For each $h \in [1,m-1]$, the following holds$:$ 
\[
D_{h}= A_{h}^{*} -A_{h} \leq \sum_{\ell=h+1}^{m} A_{\ell}=S_{h+1}.
\]
%
\end{lemma}
{\bf Proof:} Let $\varphi_{h}(e_{i})=A_{h}(e_{i})+\cdots+A_{m}(e_{i})-
A_{h}^{*}(e_{i})$. From Eqs. (\ref{eq-preserve-greedy}) and 
(\ref{eq-preserve-opt}), we have that 
\[
\varphi_{h}(e_{i}) = \sum_{\ell=h}^{m} \left\{\delta_{\ell}(e_{i})+q_{\ell}(e_{i})\right\}-
\left\{\delta_{h}^{*}(e_{i})+q_{h}^{*}(e_{i})\right\}.
\]
By induction on events $e_{i}$ for $i \geq 0$, 
we show that $\varphi_{h}(e_{i})\geq 0$. 

{\sf Base Step:} For the initial event $e_{0}$, 
it is immediate that $\delta_{h}(e_{0})=\cdots=\delta_{m}(e_{0})=0$, 
$q_{h}(e_{0})=\cdots=q_{m}(e_{0})=0$, $\delta_{h}^{*}(e_{0})=0$, and 
$q_{h}^{*}(e_{0})=0$.~This implies that $\varphi_{h}(e_{0})=0$. 

{\sf Induction Step:} For $t \geq 1$, we assume that 
$\varphi_{h}(e_{i})\geq 0$ 
for each $i \in [0,t-1]$. Let~us~consider~the case that $e_{t}$ is send event and 
the case that $e_{t}$ is arrive event. 

($e_{t}$: send event) If {\sf Opt} transmits a $v_{h}$-packet, then 
$\delta_{h}^{*}(e_{t})+q_{h}^{*}(e_{t})=\delta_{h}^{*}(e_{t-1})+1+q_{h}^{*}(e_{t-1})-1
=\delta_{h}^{*}(e_{t-1})+q_{h}^{*}(e_{t-1})$.
%
%
%
%
It is obvious that 
$\delta_{h}^{*}(e_{t})+q_{h}^{*}(e_{t})
=\delta_{h}^{*}(e_{t-1})+q_{h}^{*}(e_{t-1})$ 
if {\sf Opt} does not transmits a $v_{h}$-packet. 
For the case 
that {\sf Greedy}~transmits a $v_{r}$-packet with $r \in [h,m]$,~it~is~immediate~that $\delta_{r}(e_{t})+q_{r}(e_{t})=
\delta_{r}(e_{t-1})+1+q_{r}(e_{t-1})-1=\delta_{r}(e_{t-1})+q_{r}(e_{t-1})$~and~that $\delta_{\ell}(e_{t})+q_{\ell}(e_{t})=
\delta_{\ell}(e_{t-1})+q_{\ell}(e_{t-1})$ for each $\ell \in [h,m]\setminus\{r\}$. 
For the case that {\sf Greedy} transmits a $v_{r}$-packet~with $r \in [1,h-1]$, 
it is easy to~see~that~$\delta_{\ell}(e_{t})+q_{\ell}(e_{t})=
\delta_{\ell}(e_{t-1})+q_{\ell}(e_{t-1})$ 
for each $\ell \in [h,m]$. 
Then from the induction hypothesis, we have~that 
\begin{eqnarray*}
\varphi_{h}(e_{t}) 
& = & 
\sum_{\ell=h}^{m} \left\{\delta_{\ell}(e_{t})+q_{\ell}(e_{t})\right\} - 
\left\{\delta_{h}^{*}(e_{t}) + q_{h}(e_{t})\right\}\\
& = & \sum_{\ell=h}^{m} \left\{\delta_{\ell}(e_{t-1})+q_{\ell}(e_{t-1})\right\} - 
\left\{\delta_{h}^{*}(e_{t-1}) + q_{h}(e_{t-1})\right\}\\
& =& \varphi_{h}(e_{t-1})\geq 0. 
\end{eqnarray*}

($e_{t}$: arrive event) Notice that 
$\delta_{h}(e_{t})=\delta_{h}(e_{t-1}),\ldots,
\delta_{m}(e_{t})=\delta_{m}(e_{t-1})$ and 
$\delta_{h}^{*}(e_{t})=\delta_{h}^{*}(e_{t-1})$. 
Let us consider the following cases: 
(1) a $v_{r}$-packet with $r \in [1,h-1]$ arrives, 
(2) a $v_{r}$-packet~with $r \in [h+1,m]$ arrives, 
and 
(3) a $v_{h}$-packet arrives. 
For the case (1), 
it is immediate~that~$q_{h}(e_{t})=q_{h}(e_{t-1}),\ldots,q_{m}(e_{t})= q_{m}(e_{t-1})$ 
and $q_{h}^{*}(e_{t})=q_{h}^{*}(e_{t-1})$.  
From the induction hypothesis, it follows that 
$\varphi_{h}(e_{t})= \varphi_{h}(e_{t-1})\geq 0$. 
For the case (2), 
we have that 
$q_{r}(e_{t})\geq  q_{r}(e_{t-1})$,~$q_{\ell}(e_{t})= q_{\ell}(e_{t-1})$~for each 
$\ell \in [h,m]\setminus\{r\}$, 
and $q_{h}^{*}(e_{t})=q_{h}^{*}(e_{t-1})$. 
Thus from the induction hypothesis, it follows that $\varphi_{h}(e_{t})
\geq \varphi_{h}(e_{t-1})\geq 0$. 
For the case (3),  let us consider the following subcases: 
(3.1)~{\sf Greedy} and {\sf Opt} accept the $v_{h}$-packet, 
(3.2) {\sf Greedy} and {\sf Opt} reject the~$v_{h}$-packet, 
(3.3) {\sf Greedy} accepts the $v_{h}$-packet 
but {\sf Opt} rejects the $v_{h}$-packet, 
(3.4) {\sf Greedy} rejects the $v_{h}$-packet 
but {\sf Opt} accepts~the~$v_{h}$-packet. 
For the subcase (3.1), it is immediate that 
$q_{h}(e_{t})=q_{h}(e_{t-1})+1$, 
$q_{\ell}(e_{t})=q_{\ell}(e_{t-1})$~for~each $\ell \in [h+1,m]$, 
and $q_{h}^{*}(e_{t})=q_{h}^{*}(e_{t-1})+1$. 
From the induction hypothesis, 
it follows~that~$\varphi_{h}(e_{t})=\varphi_{h}(e_{t-1})\geq 0$. 
For the subcase (3.2),  we can show that 
$\varphi_{h}(e_{t})=\varphi_{h}(e_{t-1})\geq 0$~in~a~way~similar~to 
the subcase (3.1). 
For the subcase (3.3), we have that 
$q_{h}(e_{t})=q_{h}(e_{t-1})+1$, 
$q_{\ell}(e_{t})=q_{\ell}(e_{t-1})$~for each $\ell \in [h+1,m]$, and 
$q_{h}^{*}(e_{t})=q_{h}^{*}(e_{t-1})$. 
From the induction hypothesis, it 
follows~that~$\varphi_{h}(e_{t})=\varphi_{h}(e_{t-1})+1 \geq 0$. 
For the subcase (3.4), we have that the 
$v_{h}$-queue~of~{\sf Greedy}~is~full,~i.e., 
$q_{h}(e_{t})= B_{h}$. From the fact that $B_{h} \geq q_{h}^{*}(e_{t})$,
it is obvious that $q_{h}^{*}(e_{t}) \leq q_{h}(e_{t}) \leq 
q_{h}(e_{t})+\cdots+q_{m}(e_{t})$.~So 
from Lemma~\ref{lemma-send}~and the definition of $\varphi_{h}$, 
we have that $\varphi_{h}(e_{t}) \geq 0$. \BQED
%
\section{Competitive Ratio of the Algorithm Greedy} \label{sec-crg}
%
From Lemmas \ref{lemma-AS} and \ref{prop-acc3-eq}, it follows that 
for each $h \in [1,m-2]$, 
\begin{equation}
\sum_{\ell=h}^{m-1} D_{\ell} = \sum_{\ell=h}^{m} D_{\ell} 
\leq \sum_{\ell=h}^{m} A_{\ell}=S_{h}. \label{eq-old-ub}
\end{equation}
For each $h \in [1,m-1]$, we derive the $m-h$ upper bounds 
for $D_{h}+D_{h+1}+\cdots + D_{m-1}$~by~applying 
Eq. (\ref{eq-old-ub}) and Lemma \ref{prop-2} (see an example given in 
Appendix \ref{app-example}). 
For each $j \in [h,m-3]$, apply Lemma \ref{prop-2} to 
$D_{h},D_{h+1},\ldots,D_{j}$~and~apply Eq. (\ref{eq-old-ub}) to 
$D_{j+1}+D_{j+2}+\cdots + D_{m-1}$, i.e., 
\begin{eqnarray*}
D_{h}+D_{h+1}+\cdots + D_{m-1} & \leq & 
S_{h+1}+S_{h+1};\\
D_{h}+D_{h+1}+\cdots + D_{m-1} & \leq & 
S_{h+1}+S_{h+2}+S_{h+2};\\
   &\vdots & \\
D_{h}+D_{h+1}+\cdots + D_{m-1} & \leq & 
S_{h+1}+S_{h+2}+\cdots+ S_{j+1}+S_{j+1};\\
   &\vdots & \\
D_{h}+D_{h+1}+\cdots + D_{m-1} & \leq & 
S_{h+1}+S_{h+2}+\cdots+ S_{m-2}+S_{m-2}. 
\end{eqnarray*}
Applying Lemma \ref{prop-2} to $D_{h},D_{h+1},\ldots, D_{m-1}$, 
we have that 
\[
D_{h}+D_{h+1}+\cdots+D_{m-1} \leq 
S_{h+1}+S_{h+2}+\cdots +S_{m}, 
\]
and applying Eq. (\ref{eq-old-ub}) to $D_{h}+D_{h+1}+\cdots+D_{m-1}$, 
we also have that 
\[
D_{h}+D_{h+1}+\cdots+D_{m-1} \leq S_{h}. 
\]
Let $U_{h}$ be the minimum among $m-h$ upper bounds for 
$D_{h}+D_{h+1}+\cdots+D_{m-1}$. From~the~definition of $U_{h}$, 
it is immediate that $U_{m-1}=A_{m}$. 
For $m$ nonnegative packet values~$0<v_{1}<v_{2}<\cdots <v_{m}$, 
let ${\cal C}_{m} =\{c_{1},c_{2},\ldots,c_{m-1}\}$, where for each $i \in [1,m-1]$, 
\[
c_{i} = \frac{v_{i}+\sum_{j=1}^{i-1} 2^{j-1}v_{i-j}}{v_{i+1}+\sum_{j=1}^{i-1} 2^{j-1}v_{i-j}}. 
\]
Let $c_{m}^{*}=\max \{c_{1},c_{2},\ldots,c_{m-1}\}$. Note that $c_{m}^{*}<1$. 
The following lemmas hold for $c_{m}^{*}$ and $U_{h}$. 
\begin{lemma} \label{lemma-c}
For each $i \in [1,m-1]$, the following holds$:$
\[
\left(v_{i} + \sum_{j=1}^{i-1}2^{j-1}v_{i-j} \right) - 
c_{m}^{*}\left(v_{i+1}+\sum_{j=1}^{i-1}2^{j-1}v_{i-j} \right)\leq 0.
\]
\end{lemma}
\begin{lemma} \label{lemma-U}
For each $h \in [1,m-2]$, $U_{h}=\min \{A_{h},U_{h+1}\}+S_{h+1}$, 
where $U_{m-1}=A_{m}$. 
\end{lemma}
The proof of Lemma \ref{lemma-c} is given in Subsection 
\ref{subsec-proof-0} and 
the proof of Lemma \ref{lemma-U} is given~in~Subsection 
\ref{subsec-proof-1}. 
For each $h \in [1,m-2]$, define $\Delta_{h}$ as follows: 
\begin{eqnarray*}
\Delta_{h} & = & \left\{\left(
v_{h}+\sum_{j=1}^{h-2} 2^{j-1}v_{h-1-j} \right) 
-c_{m}^{*}\left(v_{h-1}+\sum_{j=1}^{h-2} 2^{j-1}v_{h-1-j} \right) \right\}U_{h}\\
& & ~~~~~+ 
\left\{\left(
v_{h-1}+\sum_{j=1}^{h-2} 2^{j-1}v_{h-1-j} \right) 
-c_{m}^{*}\sum_{j=1}^{h-2} 2^{j-1}v_{h-1-j} \right\}S_{h}\\
& & ~~~~~+ (v_{h+1}-v_{h})U_{h+1}+(v_{h+2}-v_{h+1})U_{h+2}+\cdots + 
(v_{m-1}-v_{m-2})U_{m-1}. 
\end{eqnarray*}
The following lemmas are crucial to analyze the competitive ratio of 
the algorithm {\sf Greedy}.  
\begin{lemma} \label{lemma-Delta-1}
For each $h \in [1,m-3]$, $\Delta_{h}\leq c_{m}^{*}v_{h}A_{h}+\Delta_{h+1}$. 
\end{lemma}
\begin{lemma} \label{lemma-Delta-2}
$\Delta_{m-2}\leq c_{m}^{*} v_{m-2}A_{m-2}+
c_{m}^{*} v_{m-1}A_{m-1}+c_{m}^{*} v_{m}A_{m}$. 
\end{lemma}
The proof of Lemma \ref{lemma-Delta-1} is given  
in Subsection \ref{subsec-proof-2} and the proof of 
Lemma \ref{lemma-Delta-2} is given~in~Subsection \ref{subsec-proof-3}. 
From Lemmas \ref{lemma-U}, \ref{lemma-Delta-1}, and 
\ref{lemma-Delta-2}, we can show the following~theorem: 
\begin{thm} \label{thm-main}
For the general $m$-valued case with class segregation, 
the online $($multi-queue$)$ buffer management algorithm 
{\sf Greedy} is $(1+c_{m}^{*})$-competitive. 
%
\end{thm}
{\bf Proof:}  For any sequence $\sigma$, it is immediate that 
\begin{eqnarray*}
\lefteqn{\frac{{\sf Opt}(\sigma)}{{\sf Greedy}(\sigma)} =  
\frac{v_{1}A_{1}^{*}+v_{2}A_{2}^{*}+\cdots+
v_{m}A_{m}^{*}}{v_{1}A_{1}+v_{2}A_{2}+\cdots+v_{m}A_{m}}}\\
& = & 1 + \frac{v_{1}(A_{1}^{*}-A_{1})+v_{2}(A_{2}^{*}-A_{2})+\cdots+
v_{m-1}(A_{m-1}^{*}-A_{m-1})+v_{m}(A_{m}^{*}-A_{m})}{v_{1}A_{1}+
v_{2}A_{2}+\cdots+v_{m}A_{m}}\nonumber\\
& = & 1 + \frac{v_{1}D_{1}+v_{2}D_{2}+\cdots+
v_{m-1}D_{m-1}}{v_{1}A_{1}+v_{2}A_{2}+\cdots+v_{m}A_{m}},
\end{eqnarray*}
where the last equality follows from Lemma \ref{prop-acc3-eq}. 
We bound $v_{1}D_{1}+v_{2}D_{2}+\cdots+v_{m-1}D_{m-1}$. 
\begin{eqnarray*}
\lefteqn{v_{1}D_{1}+v_{2}D_{2}+\cdots+v_{m-1}D_{m-1}}\\
& = & 
v_{1} (D_{1}+D_{2}+\cdots+D_{m-1})+ (v_{2}-v_{1})(D_{2}+D_{3}+\cdots+D_{m-1})\\
& & ~~~~~+ (v_{3}-v_{2})(D_{3}+D_{4}+\cdots+D_{m-1}) + \cdots + 
(v_{m-1}-v_{m-2})D_{m-1}\\
& \leq & v_{1}U_{1}+(v_{2}-v_{1})U_{2}+(v_{3}-v_{2})U_{3}+\cdots+
(v_{m-1}-v_{m-2})U_{m-1}\\
& = & \Delta_{1}, 
\end{eqnarray*}
where the inequqlity follows from 
$D_{h}+D_{h+1}+\cdots +D_{m-1}\leq U_{h}$
for each $h \in [1,m-1]$,~and~the last equality follows from 
the definition of $\Delta_{h}$. 
By the iterative use of 
Lemma \ref{lemma-Delta-1}, we have that 
\begin{eqnarray*}
\frac{{\sf Opt}(\sigma)}{{\sf Greedy}(\sigma)} & \leq & 
1 + \frac{\Delta_{1}}{v_{1}A_{1}+v_{2}A_{2}+\cdots+v_{m}A_{m}}\\
& \leq & 1 + \frac{c_{m}^{*}v_{1}A_{1}+\Delta_{2}}{v_{1}A_{1}+v_{2}A_{2}+
\cdots+v_{m}A_{m}}\\
& \leq & 1 + \frac{c_{m}^{*}v_{1}A_{1}+c_{m}^{*}v_{2}A_{2}+
\Delta_{3}}{v_{1}A_{1}+v_{2}A_{2}+\cdots+v_{m}A_{m}}\\
& \vdots & \\
& \leq & 1 + \frac{c_{m}^{*}v_{1}A_{1}+c_{m}^{*}v_{2}A_{2}+\cdots+
c_{m}^{*}v_{m-3}A_{m-3}+\Delta_{m-2}}{v_{1}A_{1}+v_{2}A_{2}+
\cdots+v_{m}A_{m}}\\ 
& \leq & 1 + \frac{c_{m}^{*}v_{1}A_{1}+c_{m}^{*}v_{2}A_{2}+\cdots+
c_{m}^{*}v_{m}A_{m}}{v_{1}A_{1}+v_{2}A_{2}+\cdots+v_{m}A_{m}}\\
& = & 1 + c_{m}^{*}\cdot\frac{v_{1}A_{1}+v_{2}A_{2}+\cdots+
v_{m}A_{m}}{v_{1}A_{1}+v_{2}A_{2}+\cdots+v_{m}A_{m}}\\
& = & 1 + c_{m}^{*}, 
\end{eqnarray*}
where all inequalities except for the first and last ones follow from Lemma 
\ref{lemma-Delta-1} and the~last~ine\-quality follows from Lemma 
\ref{lemma-Delta-2}. Thus {\sf Greedy} is $(1+c_{m}^{*})$-competitive. \BQED
%
\section{Proofs of Lemmas} \label{sec-proof-lemma}
%
\subsection{Proof of Lemma \ref{lemma-c}} \label{subsec-proof-0}
%
From the definition of $c_{i}$, it follows that for each $i \in [1,m-1]$, 
\begin{eqnarray*}
\lefteqn{\left(v_{i} + \sum_{j=1}^{i-1}2^{j-1}v_{i-j} \right) - 
c_{m}^{*}\left(v_{i+1}+\sum_{j=1}^{i-1}2^{j-1}v_{i-j} \right)}\\
& = & 
\left(v_{i+1}+\sum_{j=1}^{i-1}2^{j-1}v_{i-j} \right)
\left(\frac{v_{i} + \sum_{j=1}^{i-1}2^{j-1}v_{i-j}}{v_{i+1}+\sum_{j=1}^{i-1}2^{j-1}v_{i-j}}
- c_{m}^{*} \right)\\
& = & \left(v_{i+1}+\sum_{j=1}^{i-1}2^{j-1}v_{i-j} \right)(c_{i}-c_{m}^{*})
\leq 0, 
\end{eqnarray*}
where the inequality follows from the fact that $c_{i} \leq c_{m}^{*}$ 
for each $i \in [1,m-1]$. 
%
\subsection{Proof of Lemma \ref{lemma-U}} \label{subsec-proof-1}
%
It is obvious that $U_{m-1}=A_{m}$. 
From the definition of $U_{h}$, it follows that for each 
$i \in [1,m-2]$, 
\begin{eqnarray*}
U_{m-(i+1)} & = & \min\left\{S_{m-i}+U_{m-i},S_{m-(i+1)}\right\}
= \min\left\{S_{m-i}+U_{m-i},A_{m-(i+1)}+S_{m-i}\right\}\\
& = & \min \left\{A_{m-(i+1)},U_{m-i}\right\} + S_{m-i}. 
\end{eqnarray*}
Thus for each $h \in [1,m-2]$ we have that 
$U_{h}=\min \{A_{h},U_{h+1}\}+S_{h+1}$. 
%
\subsection{Proof of Lemma \ref{lemma-Delta-1}} \label{subsec-proof-2}
%
From Lemma \ref{lemma-U}, we have that 
$U_{h}=\min \{A_{h},U_{h+1}\}+S_{h+1}$ 
for each $h \in [1,m-2]$. Let~us~consider the following cases: 
(1) $A_{h}\leq U_{h+1}$ and (2) $A_{h} > U_{h+1}$.

For the case (1), we have that 
$U_{h}=\min \{A_{h},U_{h+1}\}+S_{h+1}=A_{h}+S_{h+1}=S_{h}$. 
So~from~the~definition of $\Delta_{h}$ and the fact that 
$U_{h}=S_{h}$, it follows that 
\begin{eqnarray*}
\Delta_{h} & = & \left\{\left(
v_{h}+\sum_{j=1}^{h-2} 2^{j-1}v_{h-1-j} \right) 
-c_{m}^{*}\left(v_{h-1}+\sum_{j=1}^{h-2} 2^{j-1}v_{h-1-j} \right) \right\}S_{h}\\
& & ~~~~~+ 
\left\{\left(
v_{h-1}+\sum_{j=1}^{h-2} 2^{j-1}v_{h-1-j} \right) 
-c_{m}^{*}\sum_{j=1}^{h-2} 2^{j-1}v_{h-1-j} \right\}S_{h}\\
& & ~~~~~+ (v_{h+1}-v_{h})U_{h+1}+(v_{h+2}-v_{h+1})U_{h+2}+\cdots + 
(v_{m-1}-v_{m-2})U_{m-1}\\
& = & \left\{\left(
v_{h}+v_{h-1}+2\sum_{j=1}^{h-2} 2^{j-1}v_{h-1-j} \right) 
-c_{m}^{*}\left(v_{h-1}+2\sum_{j=1}^{h-2} 2^{j-1}v_{h-1-j} \right) \right\}S_{h}\\
& & ~~~~~+ (v_{h+1}-v_{h})U_{h+1}+(v_{h+2}-v_{h+1})U_{h+2}+\cdots + 
(v_{m-1}-v_{m-2})U_{m-1}\\
& = & \left\{\left(
v_{h}+\sum_{j=1}^{h-1} 2^{j-1}v_{h-j} \right) 
-c_{m}^{*}\sum_{j=1}^{h-1} 2^{j-1}v_{h-j} \right\}(A_{h}+S_{h+1})\\
& & ~~~~~+ (v_{h+1}-v_{h})U_{h+1}+(v_{h+2}-v_{h+1})U_{h+2}+\cdots + 
(v_{m-1}-v_{m-2})U_{m-1}\\
& = & \left\{\left(
v_{h}+\sum_{j=1}^{h-1} 2^{j-1}v_{h-j} \right) 
-c_{m}^{*}\sum_{j=1}^{h-1} 2^{j-1}v_{h-j} \right\}A_{h}\\
& & ~~~~~+\left\{\left(
v_{h}+\sum_{j=1}^{h-1} 2^{j-1}v_{h-j} \right) 
-c_{m}^{*}\sum_{j=1}^{h-1} 2^{j-1}v_{h-j} \right\}S_{h+1}\\
& & ~~~~~+ (v_{h+1}-v_{h})U_{h+1}+(v_{h+2}-v_{h+1})U_{h+2}+\cdots + 
(v_{m-1}-v_{m-2})U_{m-1}\\
& = & c_{m}^{*}v_{h}A_{h}+\left\{\left(
v_{h}+\sum_{j=1}^{h-1} 2^{j-1}v_{h-j} \right) 
-c_{m}^{*}\left(v_{h} + \sum_{j=1}^{h-1} 2^{j-1}v_{h-j} \right)\right\}A_{h}\\
& & ~~~~~+\left\{\left(v_{h}+\sum_{j=1}^{h-1} 2^{j-1}v_{h-j} \right) 
-c_{m}^{*}\sum_{j=1}^{h-1} 2^{j-1}v_{h-j} \right\}S_{h+1}\\
& & ~~~~~+ (v_{h+1}-v_{h})U_{h+1}+(v_{h+2}-v_{h+1})U_{h+2}+\cdots + 
(v_{m-1}-v_{m-2})U_{m-1}\\
& \leq & c_{m}^{*}v_{h}A_{h}+\left\{\left(
v_{h}+\sum_{j=1}^{h-1} 2^{j-1}v_{h-j} \right) 
-c_{m}^{*}\left(v_{h} + \sum_{j=1}^{h-1} 2^{j-1}v_{h-j} \right)\right\}U_{h+1}\\
& & ~~~~~+\left\{\left(v_{h}+\sum_{j=1}^{h-1} 2^{j-1}v_{h-j} \right) 
-c_{m}^{*}\sum_{j=1}^{h-1} 2^{j-1}v_{h-j} \right\}S_{h+1}\\
& & ~~~~~+ (v_{h+1}-v_{h})U_{h+1}+(v_{h+2}-v_{h+1})U_{h+2}+\cdots + 
(v_{m-1}-v_{m-2})U_{m-1}\\
& = & c_{m}^{*}v_{h}A_{h}+\left\{\left(
v_{h+1}+\sum_{j=1}^{h-1} 2^{j-1}v_{h-j} \right) 
-c_{m}^{*}\left(v_{h} + \sum_{j=1}^{h-1} 2^{j-1}v_{h-j} \right)\right\}U_{h+1}\\
& & ~~~~~+\left\{\left(v_{h}+\sum_{j=1}^{h-1} 2^{j-1}v_{h-j} \right) 
-c_{m}^{*}\sum_{j=1}^{h-1} 2^{j-1}v_{h-j} \right\}S_{h+1}\\
& & ~~~~~+ (v_{h+2}-v_{h+1})U_{h+2}+(v_{h+3}-v_{h+2})U_{h+3}+
\cdots + (v_{m-1}-v_{m-2})U_{m-1}\\
& = & c_{m}^{*}v_{h}A_{h}+\Delta_{h+1}, 
\end{eqnarray*}
where the inequality follows from the fact that $c_{m}^{*}<1$ and 
the assumption that $A_{h}\leq U_{h+1}$. 

For the case (2), we have that $U_{h}=\min\{A_{h},U_{h+1}\}+S_{h+1}
=U_{h+1}+S_{h+1}$. 
So~from~the~definition of $\Delta_{h}$ and the fact that 
$U_{h}=U_{h+1}+S_{h+1}$, it follows that 
\begin{eqnarray*}
\Delta_{h} & = & \left\{\left(
v_{h}+\sum_{j=1}^{h-2} 2^{j-1}v_{h-1-j} \right) 
-c_{m}^{*}\left(v_{h-1}+\sum_{j=1}^{h-2} 2^{j-1}v_{h-1-j} \right) \right\}(U_{h+1}+S_{h+1})\\
& & ~~~~~+ 
\left\{\left(
v_{h-1}+\sum_{j=1}^{h-2} 2^{j-1}v_{h-1-j} \right) 
-c_{m}^{*}\sum_{j=1}^{h-2} 2^{j-1}v_{h-1-j} \right\}S_{h}\\
& & ~~~~~+ (v_{h+1}-v_{h})U_{h+1}+(v_{h+2}-v_{h+1})U_{h+2}+\cdots + 
(v_{m-1}-v_{m-2})U_{m-1}\\
& = & \left\{\left(
v_{h}+\sum_{j=1}^{h-2} 2^{j-1}v_{h-1-j} \right) 
-c_{m}^{*}\left(v_{h-1}+\sum_{j=1}^{h-2} 2^{j-1}v_{h-1-j} \right) \right\}U_{h+1}\\
& & ~~~~~+\left\{\left(
v_{h}+\sum_{j=1}^{h-2} 2^{j-1}v_{h-1-j} \right) 
-c_{m}^{*}\left(v_{h-1}+\sum_{j=1}^{h-2} 2^{j-1}v_{h-1-j} \right) \right\}S_{h+1}\\
& & ~~~~~+ 
\left\{\left(
v_{h-1}+\sum_{j=1}^{h-2} 2^{j-1}v_{h-1-j} \right) 
-c_{m}^{*}\sum_{j=1}^{h-2} 2^{j-1}v_{h-1-j} \right\}(A_{h}+S_{h+1})\\
& & ~~~~~+ (v_{h+1}-v_{h})U_{h+1}+(v_{h+2}-v_{h+1})U_{h+2}+\cdots + 
(v_{m-1}-v_{m-2})U_{m-1}\\
& = & \left\{\left(
v_{h}+\sum_{j=1}^{h-2} 2^{j-1}v_{h-1-j} \right) 
-c_{m}^{*}\left(v_{h-1}+\sum_{j=1}^{h-2} 2^{j-1}v_{h-1-j} \right) \right\}U_{h+1}\\
& & ~~~~~+\left\{\left(
v_{h}+v_{h-1}+2\sum_{j=1}^{h-2} 2^{j-1}v_{h-1-j} \right) 
-c_{m}^{*}\left(v_{h-1}+2\sum_{j=1}^{h-2} 2^{j-1}v_{h-1-j} \right) \right\}S_{h+1}\\
& & ~~~~~+ 
\left\{\left(
v_{h-1}+\sum_{j=1}^{h-2} 2^{j-1}v_{h-1-j} \right) 
-c_{m}^{*}\sum_{j=1}^{h-2} 2^{j-1}v_{h-1-j} \right\}A_{h}\\
& & ~~~~~+ (v_{h+1}-v_{h})U_{h+1}+(v_{h+2}-v_{h+1})U_{h+2}+\cdots + 
(v_{m-1}-v_{m-2})U_{m-1}\\
& = & \left\{\left(
v_{h}+\sum_{j=1}^{h-2} 2^{j-1}v_{h-1-j} \right) 
-c_{m}^{*}\left(v_{h-1}+\sum_{j=1}^{h-2} 2^{j-1}v_{h-1-j} \right) \right\}U_{h+1}\\
& & ~~~~~+\left\{\left(
v_{h}+\sum_{j=1}^{h-1} 2^{j-1}v_{h-j} \right) 
-c_{m}^{*}\sum_{j=1}^{h-1} 2^{j-1}v_{h-j} \right\}S_{h+1}\\
& & ~~~~~+ 
\left\{\left(
v_{h-1}+\sum_{j=1}^{h-2} 2^{j-1}v_{h-1-j} \right) 
-c_{m}^{*}\sum_{j=1}^{h-2} 2^{j-1}v_{h-1-j} \right\}A_{h}\\
& & ~~~~~+ (v_{h+1}-v_{h})U_{h+1}+(v_{h+2}-v_{h+1})U_{h+2}+\cdots + 
(v_{m-1}-v_{m-2})U_{m-1}\\
& = & \left\{\left(
v_{h}+\sum_{j=1}^{h-2} 2^{j-1}v_{h-1-j} \right) 
-c_{m}^{*}\left(v_{h-1}+\sum_{j=1}^{h-2} 2^{j-1}v_{h-1-j} \right) \right\}U_{h+1}\\
& & ~~~~~+\left\{\left(
v_{h}+\sum_{j=1}^{h-1} 2^{j-1}v_{h-j} \right) 
-c_{m}^{*}\sum_{j=1}^{h-1} 2^{j-1}v_{h-j} \right\}S_{h+1}\\
& & ~~~~~+ 
c_{m}^{*}v_{h}A_{h}+\left\{\left(
v_{h-1}+\sum_{j=1}^{h-2} 2^{j-1}v_{h-1-j} \right) 
-c_{m}^{*}\left(v_{h}+\sum_{j=1}^{h-2} 2^{j-1}v_{h-1-j} \right)\right\}A_{h}\\
& & ~~~~~+ (v_{h+1}-v_{h})U_{h+1}+(v_{h+2}-v_{h+1})U_{h+2}+\cdots + 
(v_{m-1}-v_{m-2})U_{m-1}\\
& \leq & \left\{\left(
v_{h}+\sum_{j=1}^{h-2} 2^{j-1}v_{h-1-j} \right) 
-c_{m}^{*}\left(v_{h-1}+\sum_{j=1}^{h-2} 2^{j-1}v_{h-1-j} \right) \right\}U_{h+1}\\
& & ~~~~~+\left\{\left(
v_{h}+\sum_{j=1}^{h-1} 2^{j-1}v_{h-j} \right) 
-c_{m}^{*}\sum_{j=1}^{h-1} 2^{j-1}v_{h-j} \right\}S_{h+1}\\
& & ~~~~~+ 
c_{m}^{*}v_{h}A_{h}+\left\{\left(
v_{h-1}+\sum_{j=1}^{h-2} 2^{j-1}v_{h-1-j} \right) 
-c_{m}^{*}\left(v_{h}+\sum_{j=1}^{h-2} 2^{j-1}v_{h-1-j} \right)\right\}U_{h+1}\\
& & ~~~~~+ (v_{h+1}-v_{h})U_{h+1}+(v_{h+2}-v_{h+1})U_{h+2}+\cdots + 
(v_{m-1}-v_{m-2})U_{m-1}\\
& = & \left\{\left(
v_{h}+v_{h-1}+2\sum_{j=1}^{h-2} 2^{j-1}v_{h-1-j} \right) 
-c_{m}^{*}\left(v_{h}+v_{h-1}+2\sum_{j=1}^{h-2} 2^{j-1}v_{h-1-j} \right) \right\}U_{h+1}\\
& & ~~~~~+c_{m}^{*}v_{h}A_{h}+\left\{\left(
v_{h}+\sum_{j=1}^{h-1} 2^{j-1}v_{h-j} \right) 
-c_{m}^{*}\sum_{j=1}^{h-1} 2^{j-1}v_{h-j} \right\}S_{h+1}\\
& & ~~~~~+ (v_{h+1}-v_{h})U_{h+1}+(v_{h+2}-v_{h+1})U_{h+2}+\cdots + 
(v_{m-1}-v_{m-2})U_{m-1}\\
& = & c_{m}^{*}v_{h}A_{h}
+\left\{\left(
v_{h}+\sum_{j=1}^{h-1} 2^{j-1}v_{h-j} \right) 
-c_{m}^{*}\left(v_{h}+\sum_{j=1}^{h-1} 2^{j-1}v_{h-j} \right) \right\}U_{h+1}\\
& & ~~~~~+\left\{\left(
v_{h}+\sum_{j=1}^{h-1} 2^{j-1}v_{h-j} \right) 
-c_{m}^{*}\sum_{j=1}^{h-1} 2^{j-1}v_{h-j} \right\}S_{h+1}\\
& & ~~~~~+ (v_{h+1}-v_{h})U_{h+1}+(v_{h+2}-v_{h+1})U_{h+2}+\cdots + 
(v_{m-1}-v_{m-2})U_{m-1}\\
& = & c_{m}^{*}v_{h}A_{h}
+\left\{\left(
v_{h+1}+\sum_{j=1}^{h-1} 2^{j-1}v_{h-j} \right) 
-c_{m}^{*}\left(v_{h}+\sum_{j=1}^{h-1} 2^{j-1}v_{h-j} \right) \right\}U_{h+1}\\
& & ~~~~~+\left\{\left(
v_{h}+\sum_{j=1}^{h-1} 2^{j-1}v_{h-j} \right) 
-c_{m}^{*}\sum_{j=1}^{h-1} 2^{j-1}v_{h-j} \right\}S_{h+1}\\
& & ~~~~~+ (v_{h+2}-v_{h+1})U_{h+2}+(v_{h+3}-v_{h+2})U_{h+3}+\cdots + 
(v_{m-1}-v_{m-2})U_{m-1}\\
& = & c_{m}^{*}v_{h}A_{h}+\Delta_{h+1}, 
\end{eqnarray*}
where the inequality follows from 
Lemma \ref{lemma-c} for $i=h-1$ and 
the assumption that $A_{h} > U_{h+1}$. 
%
\subsection{Proof of Lemma \ref{lemma-Delta-2}} \label{subsec-proof-3}
%
From Lemma \ref{lemma-U}, 
it follows that $U_{m-2}=\min \{A_{m-2},U_{m-1}\}+S_{m-1}
=\min \{A_{m-2},A_{m}\}+A_{m-1}+A_{m}$. Let us consider the following 
cases: (1) $A_{m-2} \leq A_{m}$ and (2) $A_{m-2}>A_{m}$. 

For the case (1), we have that $U_{m-2}=A_{m-2}+A_{m-1}+A_{m}=
S_{m-2}$. So from the
definition~of $\Delta_{m-2}$ and the fact that $U_{m-2}=S_{m-2}$, 
it follows that 
\begin{eqnarray*}
\Delta_{m-2} & = & \left\{\left(
v_{m-2}+\sum_{j=1}^{m-4} 2^{j-1}v_{m-3-j} \right) 
-c_{m}^{*}\left(v_{m-3}+\sum_{j=1}^{m-4} 2^{j-1}v_{m-3-j} \right)\right\}S_{m-2}\\
& & ~~~~~+ 
\left\{\left(
v_{m-3}+\sum_{j=1}^{m-4} 2^{j-1}v_{m-3-j} \right) 
-c_{m}^{*}\sum_{j=1}^{m-4} 2^{j-1}v_{m-3-j} \right\}S_{m-2}\\
& & ~~~~~+ (v_{m-1}-v_{m-2})A_{m}\\
& = & \left\{\left(
v_{m-2}+v_{m-3}+2\sum_{j=1}^{m-4} 2^{j-1}v_{m-3-j} \right) 
-c_{m}^{*}\left(v_{m-3}+2\sum_{j=1}^{m-4} 2^{j-1}v_{m-3-j} \right)\right\}S_{m-2}\\
& & ~~~~~+ (v_{m-1}-v_{m-2})A_{m}\\
& = & \left\{\left(
v_{m-2}+\sum_{j=1}^{m-3} 2^{j-1}v_{m-2-j} \right) 
-c_{m}^{*}\sum_{j=1}^{m-3} 2^{j-1}v_{m-2-j} \right\}(A_{m-2}+A_{m-1}+A_{m})\\
& & ~~~~~+ (v_{m-1}-v_{m-2})A_{m}\\
& = & c_{m}^{*}v_{m-2}A_{m-2}+c_{m}^{*}v_{m-1}A_{m-1}+c_{m}^{*}v_{m}A_{m}\\
& & ~~~~~+\left\{\left(
v_{m-2}+\sum_{j=1}^{m-3} 2^{j-1}v_{m-2-j} \right) 
-c_{m}^{*}\left(v_{m-2}+\sum_{j=1}^{m-3} 2^{j-1}v_{m-2-j} \right)\right\}A_{m-2}\\
& & ~~~~~+\left\{\left(
v_{m-2}+\sum_{j=1}^{m-3} 2^{j-1}v_{m-2-j} \right) 
-c_{m}^{*}\left(v_{m-1}+\sum_{j=1}^{m-3} 2^{j-1}v_{m-2-j} \right)\right\}A_{m-1}\\
& & ~~~~~+\left\{\left(
v_{m-2}+\sum_{j=1}^{m-3} 2^{j-1}v_{m-2-j} \right) 
-c_{m}^{*}\left(v_{m}+\sum_{j=1}^{m-3} 2^{j-1}v_{m-2-j} \right)\right\}A_{m}\\
& & ~~~~~+ (v_{m-1}-v_{m-2})A_{m}\\
& \leq & c_{m}^{*}v_{m-2}A_{m-2}+c_{m}^{*}v_{m-1}A_{m-1}+c_{m}^{*}v_{m}A_{m}\\
& & ~~~~~+\left\{\left(
v_{m-2}+\sum_{j=1}^{m-3} 2^{j-1}v_{m-2-j} \right) 
-c_{m}^{*}\left(v_{m-2}+\sum_{j=1}^{m-3} 2^{j-1}v_{m-2-j} \right)\right\}A_{m}\\
& & ~~~~~+\left\{\left(
v_{m-2}+\sum_{j=1}^{m-3} 2^{j-1}v_{m-2-j} \right) 
-c_{m}^{*}\left(v_{m-1}+\sum_{j=1}^{m-3} 2^{j-1}v_{m-2-j} \right)\right\}A_{m-1}\\
& & ~~~~~+\left\{\left(
v_{m-2}+\sum_{j=1}^{m-3} 2^{j-1}v_{m-2-j} \right) 
-c_{m}^{*}\left(v_{m}+\sum_{j=1}^{m-3} 2^{j-1}v_{m-2-j} \right)\right\}A_{m}\\
& & ~~~~~+ (v_{m-1}-v_{m-2})A_{m}\\
& = & c_{m}^{*}v_{m-2}A_{m-2}+c_{m}^{*}v_{m-1}A_{m-1}+c_{m}^{*}v_{m}A_{m}\\
& & ~~~~~+\left\{\left(
v_{m-1}+\sum_{j=1}^{m-2} 2^{j-1}v_{m-1-j} \right)
-c_{m}^{*}\left(v_{m}+\sum_{j=1}^{m-2} 2^{j-1}v_{m-1-j} \right)\right\}A_{m}\\
& & ~~~~~+\left\{\left(
v_{m-2}+\sum_{j=1}^{m-3} 2^{j-1}v_{m-2-j} \right) 
-c_{m}^{*}\left(v_{m-1}+\sum_{j=1}^{m-3} 2^{j-1}v_{m-2-j} \right)\right\}A_{m-1}\\
& = & c_{m}^{*}v_{m-2}A_{m-2}+c_{m}^{*}v_{m-1}A_{m-1}+c_{m}^{*}v_{m}A_{m}\\
& & ~~~~~+\left(v_{m}+\sum_{j=1}^{m-2} 2^{j-1}v_{m-1-j} \right)
\left(c_{m-1}-c_{m}^{*}\right) A_{m}\\
& & ~~~~~+\left(v_{m-1}+\sum_{j=1}^{m-3} 2^{j-1}v_{m-2-j} \right)
\left(c_{m-2}-c_{m}^{*}\right) A_{m-1}\\
& \leq & c_{m}^{*}v_{m-2}A_{m-2}+c_{m}^{*}v_{m-1}A_{m-1}+c_{m}^{*}v_{m}A_{m}, 
\end{eqnarray*}
where the first inequality follows from the fact $c_{m}^{*}<1$ and 
the assumption that $A_{m-2} \leq A_{m}$~and the second inequality 
follows from Lemma \ref{lemma-c} for $i=m-1$ and $i=m-2$. 

For the case (2), we have that $U_{m-2}=A_{m-1}+2A_{m}$. Then from the
definition~of $\Delta_{m-2}$ and the fact that $U_{m-2}=A_{m-1}+2A_{m}$, 
it follows that 
\begin{eqnarray*}
\Delta_{m-2} & = & \left\{\left(
v_{m-2}+\sum_{j=1}^{m-4} 2^{j-1}v_{m-3-j} \right) 
-c_{m}^{*}\left(v_{m-3}+\sum_{j=1}^{m-4} 2^{j-1}v_{m-3-j} \right)\right\}
(A_{m-1}+2A_{m})\\
& & ~~~~~+ 
\left\{\left(
v_{m-3}+\sum_{j=1}^{m-4} 2^{j-1}v_{m-3-j} \right) 
-c_{m}^{*}\sum_{j=1}^{m-4} 2^{j-1}v_{m-3-j} \right\}(A_{m-2}+A_{m-1}+A_{m})\\
& & ~~~~~+ (v_{m-1}-v_{m-2})A_{m}\\
& = & \left\{\left(
v_{m-2}+\sum_{j=1}^{m-4} 2^{j-1}v_{m-3-j} \right) 
-c_{m}^{*}\left(v_{m-3}+\sum_{j=1}^{m-4} 2^{j-1}v_{m-3-j} \right)\right\}
(A_{m-1}+A_{m})\\
& & ~~~~~+
\left\{\left(
v_{m-2}+\sum_{j=1}^{m-4} 2^{j-1}v_{m-3-j} \right) 
-c_{m}^{*}\left(v_{m-3}+\sum_{j=1}^{m-4} 2^{j-1}v_{m-3-j} \right)\right\}
A_{m}\\
& & ~~~~~+ 
\left\{\left(
v_{m-3}+\sum_{j=1}^{m-4} 2^{j-1}v_{m-3-j} \right) 
-c_{m}^{*}\sum_{j=1}^{m-4} 2^{j-1}v_{m-3-j} \right\}(A_{m-1}+A_{m})\\
& & ~~~~~+ 
\left\{\left(
v_{m-3}+\sum_{j=1}^{m-4} 2^{j-1}v_{m-3-j} \right) 
-c_{m}^{*}\sum_{j=1}^{m-4} 2^{j-1}v_{m-3-j} \right\}A_{m-2}\\
& & ~~~~~+ (v_{m-1}-v_{m-2})A_{m}\\
& = & \left\{\left(
v_{m-3}+\sum_{j=1}^{m-4} 2^{j-1}v_{m-3-j} \right) 
-c_{m}^{*}\sum_{j=1}^{m-4} 2^{j-1}v_{m-3-j} \right\}A_{m-2}\\
& & ~~~~~+
\left\{\left(
v_{m-2}+\sum_{j=1}^{m-3} 2^{j-1}v_{m-2-j} \right) 
-c_{m}^{*}\sum_{j=1}^{m-3} 2^{j-1}v_{m-2-j}\right\}
(A_{m-1}+A_{m})\\
& & ~~~~~+
\left\{\left(
v_{m-1}+\sum_{j=1}^{m-4} 2^{j-1}v_{m-3-j} \right) 
-c_{m}^{*}\left(v_{m-3}+\sum_{j=1}^{m-4} 2^{j-1}v_{m-3-j} \right)\right\}A_{m}\\
& = & c_{m}^{*}v_{m-2}A_{m-2}+\left\{\left(
v_{m-3}+\sum_{j=1}^{m-4} 2^{j-1}v_{m-3-j} \right) 
-c_{m}^{*}\left(v_{m-2}+\sum_{j=1}^{m-4} 2^{j-1}v_{m-3-j} \right)\right\}A_{m-2}\\
& & ~~~~~+
\left\{\left(
v_{m-2}+\sum_{j=1}^{m-3} 2^{j-1}v_{m-2-j} \right) 
-c_{m}^{*}\sum_{j=1}^{m-3} 2^{j-1}v_{m-2-j}\right\}A_{m-1}\\
& & ~~~~~+
\left\{\left(
v_{m-2}+\sum_{j=1}^{m-3} 2^{j-1}v_{m-2-j} \right) 
-c_{m}^{*}\sum_{j=1}^{m-3} 2^{j-1}v_{m-2-j}\right\}A_{m}\\
& & ~~~~~+
\left\{\left(
v_{m-1}+\sum_{j=1}^{m-4} 2^{j-1}v_{m-3-j} \right) 
-c_{m}^{*}\left(v_{m-3}+\sum_{j=1}^{m-4} 2^{j-1}v_{m-3-j} \right)\right\}A_{m}\\
& \leq & c_{m}^{*}v_{m-2}A_{m-2}+\left\{\left(
v_{m-3}+\sum_{j=1}^{m-4} 2^{j-1}v_{m-3-j} \right) 
-c_{m}^{*}\left(v_{m-2}+\sum_{j=1}^{m-4} 2^{j-1}v_{m-3-j} \right)\right\}A_{m}\\
& & ~~~~~+
\left\{\left(
v_{m-2}+\sum_{j=1}^{m-3} 2^{j-1}v_{m-2-j} \right) 
-c_{m}^{*}\sum_{j=1}^{m-3} 2^{j-1}v_{m-2-j}\right\}A_{m-1}\\
& & ~~~~~+
\left\{\left(
v_{m-2}+\sum_{j=1}^{m-3} 2^{j-1}v_{m-2-j} \right) 
-c_{m}^{*}\sum_{j=1}^{m-3} 2^{j-1}v_{m-2-j}\right\}A_{m}\\
& & ~~~~~+
\left\{\left(
v_{m-1}+\sum_{j=1}^{m-4} 2^{j-1}v_{m-3-j} \right) 
-c_{m}^{*}\left(v_{m-3}+\sum_{j=1}^{m-4} 2^{j-1}v_{m-3-j} \right)\right\}A_{m}\\
& = & c_{m}^{*}v_{m-2}A_{m-2}+c_{m}^{*}v_{m-1}A_{m-1}\\
& & ~~~~~+
\left\{\left(
v_{m-2}+\sum_{j=1}^{m-3} 2^{j-1}v_{m-2-j} \right) 
-c_{m}^{*}\left(v_{m-1}+\sum_{j=1}^{m-3} 2^{j-1}v_{m-2-j}\right)\right\}A_{m-1}\\
& & ~~~~~+
\left\{\left(
v_{m-2}+\sum_{j=1}^{m-3} 2^{j-1}v_{m-2-j} \right) 
-c_{m}^{*}\sum_{j=1}^{m-3} 2^{j-1}v_{m-2-j}\right\}A_{m}\\
& & ~~~~~+
\left\{\left(
v_{m-1}+\sum_{j=1}^{m-3} 2^{j-1}v_{m-2-j} \right) 
-c_{m}^{*}\left(v_{m-2}+2\sum_{j=1}^{m-3} 2^{j-1}v_{m-2-j} \right)\right\}A_{m}\\
& = & c_{m}^{*}v_{m-2}A_{m-2}+c_{m}^{*}v_{m-1}A_{m-1}\\
& & ~~~~~+
\left\{\left(
v_{m-2}+\sum_{j=1}^{m-3} 2^{j-1}v_{m-2-j} \right) 
-c_{m}^{*}\left(v_{m-1}+\sum_{j=1}^{m-3} 2^{j-1}v_{m-2-j}\right)\right\}A_{m-1}\\
& & ~~~~~+
\left\{\left(
v_{m-1}+\sum_{j=1}^{m-2} 2^{j-1}v_{m-1-j} \right) 
-c_{m}^{*}\sum_{j=1}^{m-2} 2^{j-1}v_{m-1-j}\right\}A_{m}\\
& = & c_{m}^{*}v_{m-2}A_{m-2}+c_{m}^{*}v_{m-1}A_{m-1}+c_{m}^{*}v_{m}A_{m}\\
& & ~~~~~+
\left\{\left(
v_{m-2}+\sum_{j=1}^{m-3} 2^{j-1}v_{m-2-j} \right) 
-c_{m}^{*}\left(v_{m-1}+\sum_{j=1}^{m-3} 2^{j-1}v_{m-2-j}\right)\right\}A_{m-1}\\
& & ~~~~~+
\left\{\left(
v_{m-1}+\sum_{j=1}^{m-2} 2^{j-1}v_{m-1-j} \right) 
-c_{m}^{*}\left(v_{m}+\sum_{j=1}^{m-2} 2^{j-1}v_{m-1-j}\right)\right\}A_{m}\\
& = & c_{m}^{*}v_{m-2}A_{m-2}+c_{m}^{*}v_{m-1}A_{m-1}+c_{m}^{*}v_{m}A_{m}\\
& & ~~~~~+\left(v_{m-1}+\sum_{j=1}^{m-3} 2^{j-1}v_{m-2-j}\right)
\left(c_{m-2}-c_{m}^{*}\right)A_{m-1}\\
& & ~~~~~+\left(v_{m}+\sum_{j=1}^{m-2} 2^{j-1}v_{m-1-j}\right)
\left(c_{m-1}-c_{m}^{*}\right)A_{m}\\
& \leq & c_{m}^{*}v_{m-2}A_{m-2}+c_{m}^{*}v_{m-1}A_{m-1}+c_{m}^{*}v_{m}A_{m}, 
\end{eqnarray*}
where the first inequality follows from Lemma \ref{lemma-c} for $i=m-3$
and the assumption that~$A_{m-2}>A_{m}$ and the second inequality 
follows from Lemma \ref{lemma-c} for $i=m-2$ and $i=m-1$.
%
\section{Concluding Remarks} \label{remarks}
%
In this paper, we have focused on multi-queue buffer management 
with class segregation.~In~particular, we have dealt with the general 
$m$-valued case (packet values are $0<v_{1}<v_{2}<\cdots <v_{m}$)  
and 
have shown (in Theorem \ref{thm-main}) that {\sf Greedy} is $(1+c_{m}^{*})$-competitive, where 
$c_{m}^{*} = \max_{i} c_{i}$ and 
\[
c_{i} = \frac{v_{i}+\sum_{j=1}^{i-1}2^{j-1}v_{i-j}}{v_{i+1}+\sum_{j=1}^{i-1}2^{j-1}v_{i-j}}
\]
for each $ i \in [1,m-1]$. 
Al-Bawani and Souza \cite[Theorem 3.1]{ABS} showed that for the 
general-valued case,~the~competi\-tive ratio of any deterministic online 
algorithm is at least 
\[
2 - \frac{v_{m}}{v_{m}+v_{m-1}+\cdots+v_{1}}
=1+\frac{v_{m-1}+\cdots+v_{1}}{v_{m}+v_{m-1}+\cdots+v_{1}}. 
\]
For the general $m$-valued case, it is immediate that 
\[
1 + c_{m}^{*} \geq 1 + c_{m-1} >  
1+\frac{v_{m-1}+\cdots+v_{1}}{v_{m}+v_{m-1}+\cdots+v_{1}}, 
\]
which implies that there might be gap between 
the lower and upper bounds for the competitive ratio of 
the algorithm {\sf Greedy}. To precisely capture the inherent nature of 
the multi-queue~buffer management 
with class segregation, the following problems are left to solve.
\begin{namelist}{~~~(2)}
\item[(1)] Design an efficient 
online $($multi-queue$)$ buffer management algorithm 
for the general~$m$-valued case to improve the upper bound of 
the competitive ratio.
\item[(2)] Improve the lower bound of the competitive ratio 
for the general $m$-valued case. 
\end{namelist}
%
%

%
\appendix
%
\section{An Example} \label{app-example}
%
Let $m=5$. From Lemma \ref{prop-2} and Eq. (\ref{eq-old-ub}), 
we have the following inequalities. 
\begin{eqnarray}
\begin{array}{ccccccc}
\makebox[0.25cm]{$D_{1}$} & 
\makebox[0.25cm]{} & 
\makebox[0.25cm]{} & 
\makebox[0.25cm]{} & 
\makebox[0.25cm]{} & 
\makebox[0.25cm]{} & 
\makebox[0.25cm]{} 
\end{array} & \leq & 
\begin{array}{cccccccccc}
\makebox[0.5cm]{}  & 
\makebox[0.25cm]{}  &  
\makebox[0.25cm]{$S_{2}$} & 
\makebox[0.25cm]{} & 
\makebox[0.25cm]{} & 
\makebox[0.25cm]{} & 
\makebox[0.25cm]{} & 
\makebox[0.25cm]{} & 
\makebox[0.25cm]{}
\end{array} \label{eq-m=5-1}\\
\begin{array}{ccccccc}
\makebox[0.25cm]{} & 
\makebox[0.25cm]{} & 
\makebox[0.25cm]{$D_{2}$} & 
\makebox[0.25cm]{} & 
\makebox[0.25cm]{} & 
\makebox[0.25cm]{} & 
\makebox[0.25cm]{}
\end{array} & \leq & 
\begin{array}{ccccccccc}
\makebox[0.5cm]{}  & 
\makebox[0.25cm]{}  &  
\makebox[0.25cm]{} & 
\makebox[0.25cm]{} & 
\makebox[0.25cm]{$S_{3}$} & 
\makebox[0.25cm]{} & 
\makebox[0.25cm]{} & 
\makebox[0.25cm]{} & 
\makebox[0.25cm]{}
\end{array} \label{eq-m=5-2}\\
\begin{array}{ccccccc}
\makebox[0.25cm]{} & 
\makebox[0.25cm]{} & 
\makebox[0.25cm]{} & 
\makebox[0.25cm]{} & 
\makebox[0.25cm]{$D_{3}$} & 
\makebox[0.25cm]{} & 
\makebox[0.25cm]{} 
\end{array} & \leq & 
\begin{array}{ccccccccc}
\makebox[0.5cm]{}  & 
\makebox[0.25cm]{}  &  
\makebox[0.25cm]{} & 
\makebox[0.25cm]{} & 
\makebox[0.25cm]{} & 
\makebox[0.25cm]{} & 
\makebox[0.25cm]{$S_{4}$} & 
\makebox[0.25cm]{} & 
\makebox[0.25cm]{} 
\end{array} \label{eq-m=5-3}\\
\begin{array}{ccccccc}
\makebox[0.25cm]{} & 
\makebox[0.25cm]{} & 
\makebox[0.25cm]{} & 
\makebox[0.25cm]{} & 
\makebox[0.25cm]{} & 
\makebox[0.25cm]{} & 
\makebox[0.25cm]{$D_{4}$}
\end{array} & \leq & 
\begin{array}{ccccccccc}
\makebox[0.5cm]{}  & 
\makebox[0.25cm]{}  &  
\makebox[0.25cm]{} & 
\makebox[0.25cm]{} & 
\makebox[0.25cm]{} & 
\makebox[0.25cm]{} & 
\makebox[0.25cm]{} & 
\makebox[0.25cm]{} & 
\makebox[0.25cm]{$S_{5}$}
\end{array} \label{eq-m=5-4}\\
\begin{array}{ccccccc}
\makebox[0.25cm]{$D_{1}$} & 
\makebox[0.25cm]{$+$} & 
\makebox[0.25cm]{$D_{2}$} & 
\makebox[0.25cm]{$+$} & 
\makebox[0.25cm]{$D_{3}$} & 
\makebox[0.25cm]{$+$} & 
\makebox[0.25cm]{$D_{4}$}
\end{array} & \leq & 
\begin{array}{ccccccccc}
\makebox[0.5cm]{$S_{1}$}  & 
\makebox[0.25cm]{}  &  
\makebox[0.25cm]{} & 
\makebox[0.25cm]{} & 
\makebox[0.25cm]{} & 
\makebox[0.25cm]{} & 
\makebox[0.25cm]{} & 
\makebox[0.25cm]{} & 
\makebox[0.25cm]{}
\end{array} \label{eq-m=5-5}\\
\begin{array}{ccccccc}
\makebox[0.25cm]{} & 
\makebox[0.25cm]{} & 
\makebox[0.25cm]{$D_{2}$} & 
\makebox[0.25cm]{$+$} & 
\makebox[0.25cm]{$D_{3}$} & 
\makebox[0.25cm]{$+$} & 
\makebox[0.25cm]{$D_{4}$}
\end{array} & \leq & 
\begin{array}{ccccccccc}
\makebox[0.5cm]{}  & 
\makebox[0.25cm]{}  &  
\makebox[0.25cm]{$S_{2}$} & 
\makebox[0.25cm]{} & 
\makebox[0.25cm]{} & 
\makebox[0.25cm]{} & 
\makebox[0.25cm]{} & 
\makebox[0.25cm]{} & 
\makebox[0.25cm]{}
\end{array} \label{eq-m=5-6}\\
\begin{array}{ccccccc}
\makebox[0.25cm]{} & 
\makebox[0.25cm]{} & 
\makebox[0.25cm]{} & 
\makebox[0.25cm]{} & 
\makebox[0.25cm]{$D_{3}$} & 
\makebox[0.25cm]{$+$} & 
\makebox[0.25cm]{$D_{4}$}
\end{array} & \leq & 
\begin{array}{ccccccccc}
\makebox[0.5cm]{}  & 
\makebox[0.25cm]{}  &  
\makebox[0.25cm]{} & 
\makebox[0.25cm]{} & 
\makebox[0.25cm]{$S_{3}$} & 
\makebox[0.25cm]{} & 
\makebox[0.25cm]{} & 
\makebox[0.25cm]{} & 
\makebox[0.25cm]{}
\end{array} \label{eq-m=5-7}
\end{eqnarray}
For $D_{1}+D_{2}+D_{3}+D_{4}$, we have the following 
four upper bounds:
\[
\begin{array}{rcll}
D_{1}+D_{2}+D_{3}+D_{4} & \leq & S_{2}+S_{2}; & 
\mbox{~~~from Ineqs. (\ref{eq-m=5-1}) and (\ref{eq-m=5-6})}\\
D_{1}+D_{2}+D_{3}+D_{4} & \leq & S_{2}+S_{3}+S_{3}; & 
\mbox{~~~from Ineqs. (\ref{eq-m=5-1}), (\ref{eq-m=5-2}), and (\ref{eq-m=5-7})}\\
D_{1}+D_{2}+D_{3}+D_{4} & \leq & S_{2}+S_{3}+S_{4}+S_{5};  &
\mbox{~~~from Ineqs. (\ref{eq-m=5-1}), (\ref{eq-m=5-2}), 
(\ref{eq-m=5-3}), and (\ref{eq-m=5-4})}\\
D_{1}+D_{2}+D_{3}+D_{4} & \leq & S_{1}. & 
\mbox{~~~from Ineq. (\ref{eq-m=5-5})}
\end{array}
\]
For $D_{2}+D_{3}+D_{4}$, we also have the following three upper bounds. 
\[
\begin{array}{rcll}
D_{2}+D_{3}+D_{4} & \leq & S_{3}+S_{3}; & 
\mbox{~~~from Ineqs. (\ref{eq-m=5-2}) and (\ref{eq-m=5-7})}\\
D_{2}+D_{3}+D_{4} & \leq & S_{3}+S_{4}+S_{5}; & 
\mbox{~~~from Ineqs. (\ref{eq-m=5-2}), (\ref{eq-m=5-3}), 
and (\ref{eq-m=5-4})}\\
D_{2}+D_{3}+D_{4} & \leq & S_{2}. & 
\mbox{~~~from Ineq. (\ref{eq-m=5-6})}
\end{array}
\]
For $D_{3}+D_{4}$, we also have the following two upper bounds. 
\[
\begin{array}{rcll}
D_{3}+D_{4} & \leq & S_{4}+S_{5}; & 
\mbox{~~~from Ineqs. (\ref{eq-m=5-3}) and (\ref{eq-m=5-4})}\\
D_{3}+D_{4} & \leq & S_{3}, & 
\mbox{~~~from Ineq. (\ref{eq-m=5-7})}
\end{array}
\]
and we finally have that $D_{4}\leq S_{5}$. 

\end{document}